\documentclass[useAMS,usenatbib]{mn2e}
\usepackage{journals}
\usepackage{graphicx}
\usepackage{color}
\usepackage{times}

\newcommand{\kms} {\hbox{${\rm km}\:{\rm s}^{-1}$}}

\title[Chemical abundances and kinematics of 257 G-, K-type field giants.]
{Chemical abundances and kinematics of 257 G-, K-type field giants. Setting a base for further analysis of giant-planet properties 
orbiting evolved stars\thanks{Based on observations collected at the Paranal Observatory, ESO (Chile)
with the Ultra-violet and Visible Echelle Spectrograph (UVES) of the VLT,
under programmes 085.C-0062 and 086.C-0098.}}
\author[V.~Zh.~Adibekyan et al.]{V.~Zh.~Adibekyan,$^{1}$\thanks{E-mail:
Vardan.Adibekyan@astro.up.pt}
L.~ Benamati,$^{1,2}$
N.~C.~Santos,$^{1,2}$
S.~Alves,$^{3,4}$
C.~Lovis,$^{5}$
\newauthor
S.~Udry,$^{5}$
G.~Israelian,$^{6,7}$
S.~G.~Sousa,$^{1}$
M.~Tsantaki,$^{1,2}$
A.~Mortier,$^{8}$
A.~Sozzetti,$^{9}$ 
\newauthor
and J.~R.~De Medeiros$^{10}$
\\
$^{1}$Instituto de Astrof\'{\i}ísica e Ci\^{e}ncia do Espa\c{c}o, Universidade do Porto, CAUP, Rua das Estrelas, PT4150-762 Porto, Portugal \\
$^{2}$ Departamento de F\'{\i}ísica e Astronomia, Faculdade de Ci\^{e}ncias da Universidade do Porto, Portugal \\
$^{3}$ Instituto de Astrof\'isica, Pontificia Universidad Cat\'olica de Chile, Av. Vicu\~na Mackenna 4860, 782-0436, Macul, Santiago, Chile \\
$^{4}$ CAPES Foundation, Ministry of Education of Brazil, Bras\'ilia / DF, Brazil \\
$^{5}$ Observatoire de Gen\`eve, Université de Genève, 51 ch. des Maillettes, 1290, Sauverny, Switzerland \\
$^{6}$ Instituto de Astrof\'isica de Canarias, 38200 La Laguna, Tenerife, Spain  \\
$^{7}$ Departamento de Astrof\'isica, Universidad de La Laguna, 38206 La Laguna, Tenerife, Spain \\
$^{8}$ SUPA, School of Physics and Astronomy, University of St Andrews, St Andrews KY16 9SS, UK \\
$^{9}$ INAF -- Osservatorio Astrofisico di Torino, 10025 Pino Torinese, Italy \\
$^{10}$ Departamento de F\'isica Te\'orica e Experimental, Universidade Federal do Rio Grande do Norte, Campus Universit\'{a}rio Lagoa Nova, 
59072-970, Natal, RN, Brasil}

\begin{document}

\date{Accepted ... Received ...; in original form ...}

\pagerange{\pageref{firstpage}--\pageref{lastpage}} \pubyear{2002}

\maketitle

\label{firstpage}

\begin{abstract}
We performed a uniform and detailed abundance analysis of 12 refractory elements (Na, Mg, Al, Si, Ca, Ti, Cr, Ni, Co, Sc, Mn, and V) 
for a sample of 257 G- and K-type evolved stars from the CORALIE planet search program. To date, only one of these stars is known to 
harbor a planetary companion. We aimed to characterize this large sample of evolved stars in terms of chemical abundances and kinematics,
thus setting a solid base for further analysis of planetary properties around giant stars. This sample, being homogeneously 
analyzed, can be used as a comparison sample for other planet-related studies, as well as for different type of studies related to 
stellar and Galaxy astrophysics.
The abundances of the chemical elements were determined using an LTE abundance analysis relative to the Sun, with the 
spectral synthesis code MOOG and a grid of Kurucz ATLAS9 atmospheres. To separate the Galactic stellar populations
both a purely kinematical approach and a chemical method were applied.
We confirm the overabundance of Na in giant stars compared to the field
FGK dwarfs. This enhancement might have a stellar evolutionary character, but departures from LTE  may also produce a 
similar enhancement. Our chemical separation of stellar populations also suggests a ``gap'' in metallicity between
the thick-disk and high-$\alpha$ metal-rich stars, as previously observed in dwarfs sample from HARPS.
The present sample, as most of the giant star samples, also suffers from the B -- V colour cut-off, which excludes
low-{$\log g$} stars with high metallicities, and high-{$\log g$} star with low-[Fe/H]. For future studies of planet occurrence dependence on stellar metallicity around these evolved stars
we suggest to use a sub-sample of stars in a ``cut-rectangle'' in the {$\log g$} -- [Fe/H] diagram to overcome the 
aforementioned issue.
\end{abstract}


\section{Introduction}

The precise chemical and kinematic characterization of intermediate-mass, evolved stars is very important for 
different fields of both Galactic and stellar astronomy, and the emerging field of planetary sciences.

Many studies observed significant differences in chemical abundances between main sequence dwarf and evolved stars 
\citep[e.g.,][]{Friel-03, Jacobson-07, Villanova-09, Santrich-13}. While these differences for some elements might by astrophysical, 
having a stellar evolutionary character \citep[e.g.,][for sodium]{Tautvaisiene-05}, several authors however suggested that
the differences may arise also in the analysis, being dependent on the particular method and line-list used \citep[e.g.,][]{Santos-09}.
Along the same line, one should consider also non Local Thermodynamic Equilibrium (non-LTE) effects which are stronger
for giants than for dwarfs and may have a strong influence on the analysis \citep[e.g.,][]{Bergemann-13, Bergemann-14, Alexeeva-14}. 

Understanding the mentioned issues, will not only allow to improve  of stellar atmosphere models, but also will have 
very important implications in several fields of astrophysics. 
For instance, it would help us  shed light on the statistical and evolutionary properties of planetary systems around giant stars, 
e.g., on the possible absence of the correlation between stellar metallicity and formation efficiency of giant planets
\citep[e.g.,][]{Pasquini-07, Takeda-08, Ghezzi-10, Mortier-13, Maldonado-13, Jofre-15}%
\footnote{Indeed, \citet{Reffert-14} claims a strong evidence for a planet-metallicity correlation 
for giant planet host stars.}
which was found for main sequence dwarf stars \citep[e.g.,][]{Gonzalez-97, Santos-01, Santos-04, Fischer-05, Johnson-10, Sousa-11, 
Mortier-13b}.

Several explanations have been suggested for the  aforementioned lack of metallicity enhancement for giant stars hosting 
a giant planet. Higher stellar mass of giants may compensate the lack of metals \citep[e.g.,][]{Ghezzi-10}; 
possible spectroscopic analysis issues in giant stars \citep[e.g.,][]{Hekker-07, Santos-09}; selection biases in giant star samples 
\citep{Mortier-13}. However, one should note that some studies reported an enhanced metallicity of giant stars with planets, but with small
samples of planet hosts \citep[][]{Hekker-07,Quirrenbach-11}. We refer the reader to \citet[][and reference therein]{Alves-15} 
for more detailed review on the topic.

In this paper we  focus on the chemical and kinematic properties of a sample of 257 field giant stars which 
are observed within the context of the CORALIE extrasolar planet search program. The main characteristics of the sample along with
the homogeneously derived stellar atmospheric parameters are presented in a parallel paper \citep{Alves-15}. The uniform chemical
analysis of these giant stars is very important to explore the specific chemical requirements for the formation and evolution of planetary systems 
around them.
The paper is organized as follows: in Sect. 2, we briefly introduce the sample used in this work. The method of the chemical 
abundance determination and analysis will be explained in Sect. 3. The distinction of different Galactic stellar populations
and kinematic properties of the stars are presented in Sect. 4. Then, after discussing the metallicity distribution of the stars
in Sect. 5, we summarize our main results in Sect. 6.

\section{Sample description and stellar parameters}

Our sample comprises of 257 G- and K-type evolved  stars that are being surveyed for planets in the context of the CORALIE \citep{Udry-00} extrasolar planet
search program. High-resolution and high signal-to-noise (S/N) spectra were obtained using the UVES spectrograph.
Precise stellar parameters for the entire sample were determined in \citet{Alves-15} by using the same spectra as we did for this study. 
The spectroscopic stellar parameters and metallicities were derived by imposing excitation and ionization equilibrium.
The spectroscopic analysis was completed assuming LTE with a grid of Kurucz atmosphere 
models \citep{Kurucz-93}, and the 2002 version of the MOOG%
\footnote{ The source code of MOOG can be downloaded at \texttt{http://www.as.utexas.edu/$\sim$chris/moog.html}%
} radiative transfer code  \citep{Sneden-73}. We refer the reader to \citet{Alves-15} and \citet{Sousa-14} for details. 

\citet{Alves-15} derived the atmospheric parameters by using three different line-lists of FeI and  FeII 
\citep{Hekker-07, Sousa-08, Tsantaki-13}. Whilst showing that the use of different line-lists gives compatible results, 
the parameters derived following \citet{Tsantaki-13} were adopted, so we also do for the rest of the present paper.

The stars in the sample have effective temperatures 4700 $\lesssim$ \emph{$T{}_{\mathrm{eff}}$} $\lesssim$ 5600 \emph{K},
surface gravities 2.2 $\lesssim$ {$\log g$} $\lesssim$ 3.7 dex, microturbulence 1 $\lesssim$ \emph{$\xi{}_{\mathrm{t}}$} $\lesssim$ 3.2 \kms
and they lie in the metallicity range of -0.75 $\lesssim$ {[}Fe/H{]} $\lesssim$ 0.3 dex.

\section{Chemical abundances}

For the abundance derivation we closely followed the method described in \citet{Adibekyan-12b}.

\begin{figure*}
\begin{center}$
\begin{array}{lll}
\includegraphics[angle=0,width=0.33\linewidth]{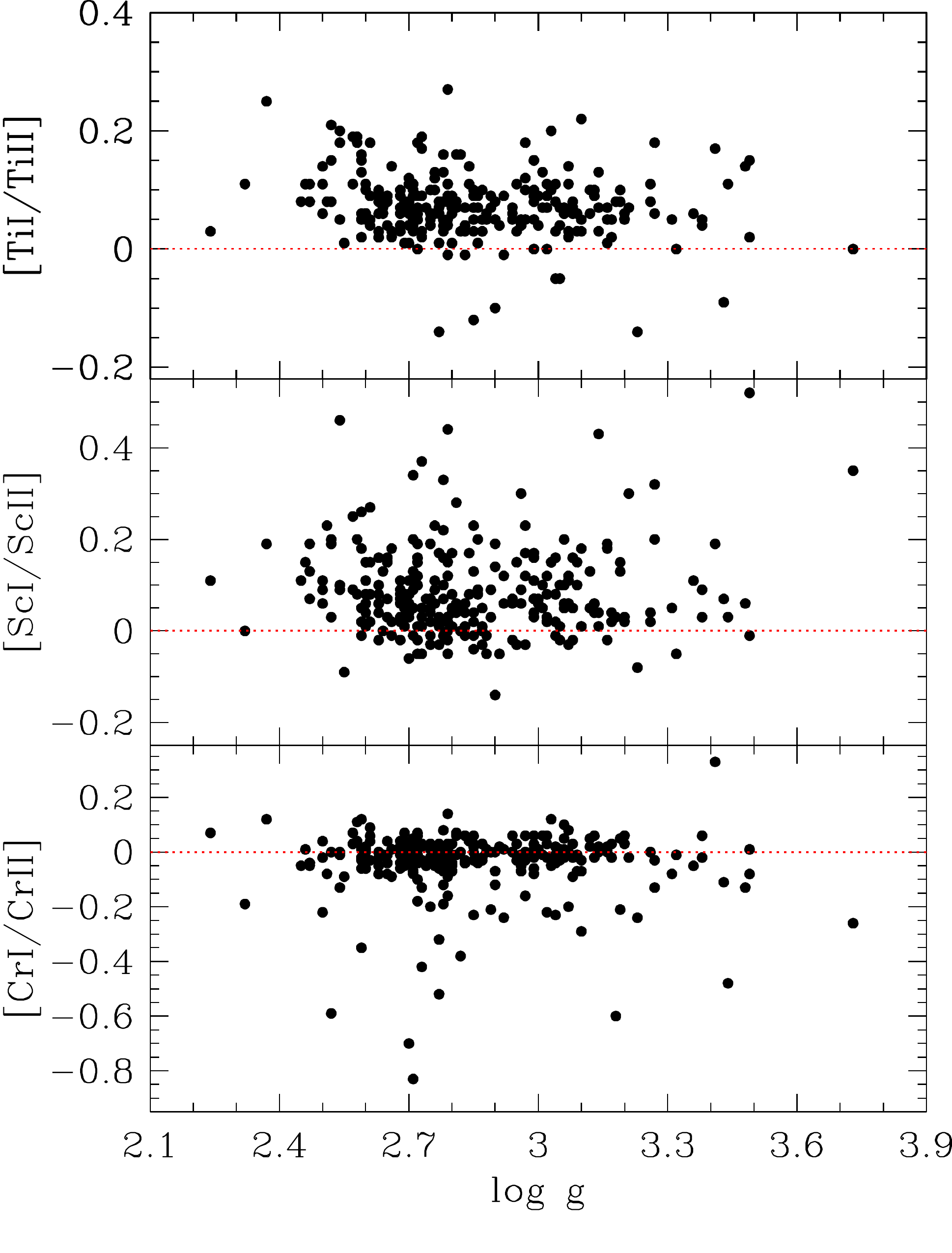}
\includegraphics[angle=0,width=0.32\linewidth]{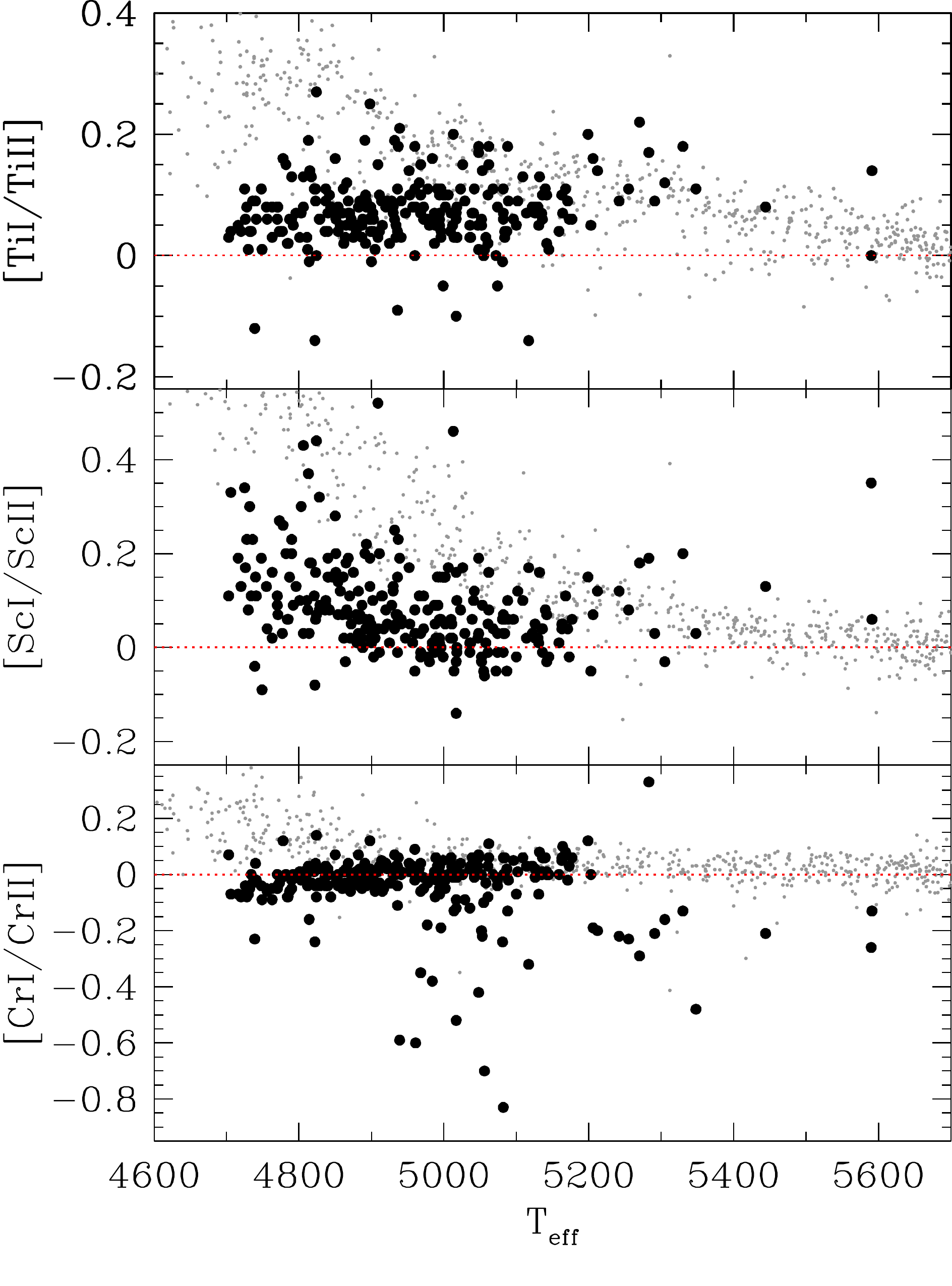} \hspace{0.1cm}
\includegraphics[angle=0,width=0.32\linewidth]{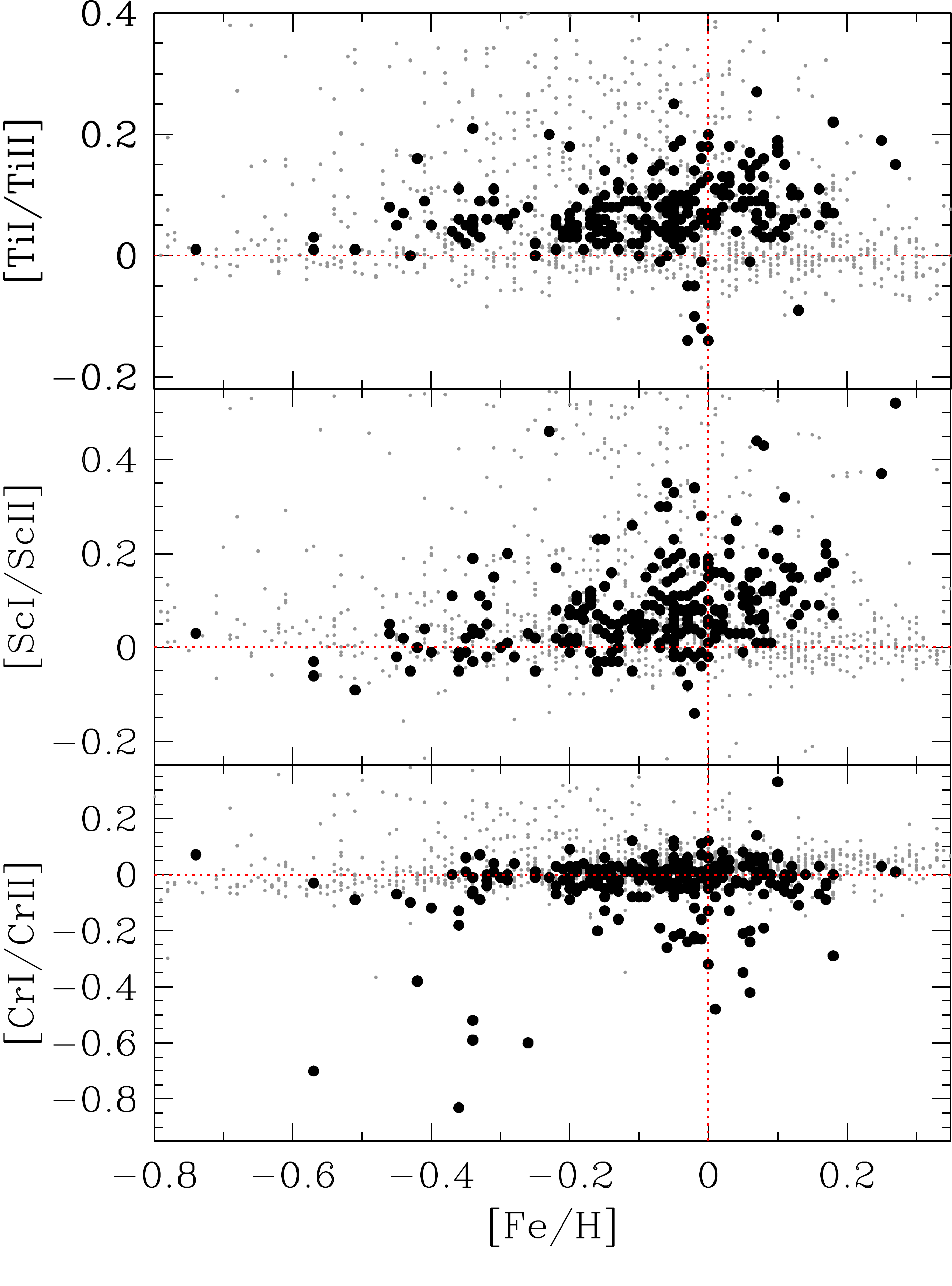}
\end{array}$
\end{center}
\vspace{-0.7cm}
\caption{[CrI/CrII], [ScI/ScII], and [TiI/TiII] as a function of atmospheric parameters for our sample of evolved stars (black points) and for
the sample of FGK dwarf stars from \citet{Adibekyan-12b} (grey dots). }
\label{fig_x1x2_param}
\end{figure*}

\begin{figure*}
\begin{center}
\begin{tabular}{c}
\includegraphics[angle=0,width=0.8\linewidth]{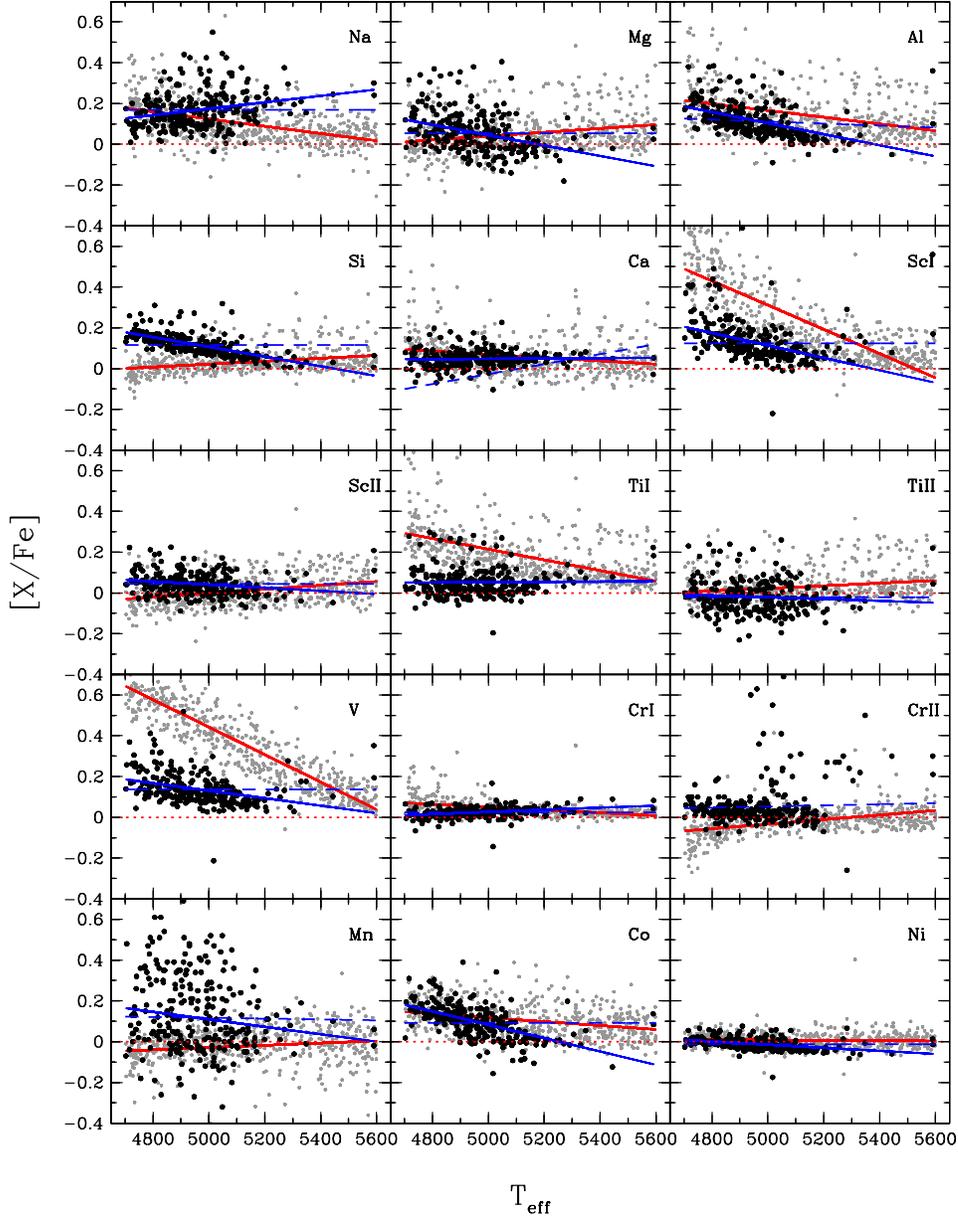}
\end{tabular}
\end{center}
\vspace{-1.3cm}
\caption{[X/Fe] vs. \emph{$T{}_{\mathrm{eff}}$} plots. 
The black dots represent the stars of the sample and the gray small dots represent stars from \citet{Adibekyan-12b}. 
The blue and red solid lines depict the linear fits of the current data
and the data from \citet{Adibekyan-12b}, respectively. The blue dashed line is the fit of our data after correcting for the trend with
\emph{$T{}_{\mathrm{eff}}$}.
Each element is identified in the \emph{upper right corner} of the respective plot.}
\label{fig_xfe_teff}
\end{figure*}

\begin{table}
\centering
\caption{The slope, correlation coefficient, and the significance of the [X/Fe] linear trends with the \emph{$T{}_{\mathrm{eff}}$}.}
\label{table-2}
\begin{tabular}{lcccc}
\hline 
\hline
\noalign{\vskip0.01\columnwidth} 
Elem & slope & R$^{2}$ & N & z-score\tabularnewline
\hline 
NaI & 1.62$\pm$0.36 $\times$ 10$^{-4}$ & 0.72 $\times$ 10$^{-1}$ & 256 & 4.3 \tabularnewline
MgI & -2.51$\pm$0.38 $\times$ 10$^{-4}$ & 0.14 $\times$ 10$^{0}$  & 257 & 6.1 \tabularnewline
AlI & -2.69$\pm$0.29 $\times$ 10$^{-4}$ & 0.24 $\times$ 10$^{0}$  & 257 & 7.9 \tabularnewline
SiI & -2.39$\pm$0.16 $\times$ 10$^{-4}$ & 0.45 $\times$ 10$^{0}$  & 257 & 11.0 \tabularnewline
CaI & 0.61$\pm$1.80 $\times$ 10$^{-5}$ & 0.44 $\times$ 10$^{-3}$  & 257 & 0.3 \tabularnewline
ScI & -3.00$\pm$0.34 $\times$ 10$^{-4}$ & 0.23 $\times$ 10$^{0}$  & 255 & 7.6 \tabularnewline
ScII & -0.78$\pm$0.23 $\times$ 10$^{-4}$ & 0.43 $\times$ 10$^{-1}$  & 257 & 3.2 \tabularnewline
TiI & 0.81$\pm$2.94 $\times$ 10$^{-5}$ & 0.29 $\times$ 10$^{-3}$  & 257 & 0.2 \tabularnewline
TiII & -0.44$\pm$0.34 $\times$ 10$^{-4}$ & 0.64 $\times$ 10$^{-2}$  & 257 & 1.2 \tabularnewline
VI & -1.79$\pm$0.29 $\times$ 10$^{-4}$ & 0.13 $\times$ 10$^{0}$  & 256 & 5.9 \tabularnewline
CrI & 0.56$\pm$0.10 $\times$ 10$^{-4}$ & 0.98 $\times$ 10$^{-1}$  & 257 & 4.9 \tabularnewline
CrII & 1.33$\pm$0.34 $\times$ 10$^{-4}$ & 0.57 $\times$ 10$^{-1}$  & 246 & 3.6 \tabularnewline
MnI & 1.33$\pm$0.34 $\times$ 10$^{-4}$ & 0.57 $\times$ 10$^{-1}$  & 247 & 2.3 \tabularnewline
CoI & -0.32$\pm$0.03 $\times$ 10$^{-3}$ & 0.33 $\times$ 10$^{0}$  & 257 & 9.2 \tabularnewline
NiI & -0.78$\pm$0.11 $\times$ 10$^{-4}$ & 0.16 $\times$ 10$^{0}$  & 257 & 6.3\tabularnewline
\hline 
\end{tabular}
\end{table}

\subsection{Selection of the lines and abundance derivation}

The initial line-list and the atomic data were taken from \citet{Adibekyan-12b} and \citet{Neves-09}. 
\citet{Neves-09} provieded the astrophysical (callibrated) oscillator strength and solar equivalent widths of the lines.
Since the spectra of cool evolved stars are more line crowded (which cause
strong blending) compared to their unevolved hotter counterparts, we aimed to carefully select a subset of unblended lines from \citet{Adibekyan-12b}. 
For this purpose as a reference we used a very high S/N and high resolution 
archival spectrum of the K-type giant Arcturus observed with the NARVAL spectrograph \citep{Mortier-13}. 
We measured the equivalent widths (EWs) of the selected lines both manually, using a Gaussian fitting procedure within the IRAF%
\footnote{IRAF is distributed by National Optical Astronomy Observatories, operated by the Association of Universities for
Research in Astronomy, Inc., under contract with the National Science Foundation, USA.}
\texttt{splot} task, and automatically, by using the ARES%
\footnote{The ARES code can be downloaded at http://www.astro.up.pt/$\sim$sousasag/ares%
} code \citep{Sousa-07}. We calculated the mean relative difference ((EW$_{ARES}$ - EW$_{IRAF}$)/ EW$_{IRAF}$) and standard deviation of the relative difference
of the EW measurements and applied 2$\sigma$-clipping. We repeated this procedure a second time after the outliers were excluded. 
Finally, 118 lines out of 164 were left that show a relative difference in EW of less than 15\%. These lines were once again checked by eye within IRAF to make sure that they are not blended and
hence the correspondence between the EW measurements is not by chance%
\footnote{The line-list is available at the CDS%
}. 

After selecting the isolated lines, the abundances for 12 elements (Na, Mg, Al, Si, Ca, Ti, Cr, Ni, Co, Sc, Mn, and V) were determined using a LTE 
analysis relative to the Sun with the 2010 version of the MOOG \citep{Sneden-73} and a grid of Kurucz ATLAS9 plane-parallel model 
atmospheres with no $\alpha$-enhancement.
The reference abundances used in the abundance analysis were taken from \citet{Anders-89}. 
We note that our analysis is differential and the source of the reference abundances is not crucial.
For the automatic EW measurements we used ARES for which
the input parameters were the same as in \citet{Sousa-08} and the
\textit{rejt} parameter is calculated following the procedure discussed in \citet{Mortier-13}. The EWs of all the lines used in the derivation 
of the abundances for all studied stars is available at CDS.

The final abundance for each star and element was calculated to be the average value of the abundances given by all lines detected
in a given star and element. Individual lines for a given star and element with a line dispersion more than a factor of
two higher than the {\it rms} were excluded. In this way we avoided the errors caused by bad pixels, cosmic rays, or other unknown effects.
A sample of our results for three stars is presented in Table~\ref{table-abundance} and the complete results are available at the
CDS.

\subsection{Uncertainties}

Since the abundances were determined via the measurement of EWs and using already determined stellar parameters, the 
errors might come from the EW measurements, from the errors in the atomic parameters, and from the uncertainties of the atmospheric parameters 
that were used to make an atmosphere model. In addition to the above-mentioned errors, one should add systematic errors 
that can occur due to NLTE or granulation (3D) effects. To minimize the errors, it is very important to use high-quality data and  
as many lines as possible for each element.

We followed \citet{Adibekyan-12b} for the calculation of the errors. In short, we first varied the model parameters by an 
amount of their one-sigma errors available for each star and calculated the abundance differences between the values 
obtained with and without varying the parameter. Then we evaluated the errors in the abundances of all elements [X/H], adding quadratically 
the line-to-line scatter errors  and errors induced by uncertainties in the model atmosphere parameters. In cases when only one line used to derive the 
abundances a typical 0.1 dex error for line-to-line scatter was assumed. 
For our sample stars the errors induced by uncertainties in the parameters of model atmosphere varies from about 0.02 dex (for SiI) to 
$\approx$ 0.06 dex (for VI) and in general are smaller than the line-to-line scatter errors. The final errors for the studied elements 
are smallest for AlI ($\approx$ 0.04 dex) and largest for VI ($\approx$ 0.14dex).

\begin{table*}
\begin{center}
\caption{Sample table of the derived abundances of the elements, rms, total error, and number of measured lines for each star.}
\label{table-abundance}
\begin{tabular}{c|cccccc|cccc|c}
\hline
\hline
\noalign{\vskip0.01\columnwidth}
Star & ... & {[}SiI/H{]} & rms & err & {[}SiI/H{]}$_{corr}^{*}$ & n & {[}CaI/H{]} & rms & err & n & ...\tabularnewline
\hline 
... & ... & ... & ... & ... & ... & ... & ... & ... & ... & ... & ... \tabularnewline
HD47001  &  ...  & -0.20 & 0.09 & 0.09 & -0.26 & 14 & -0.25 & 0.05 & 0.06 & 11 &   ...\tabularnewline
HD73898  &  ...  & -0.30 & 0.03 & 0.03 & -0.28 & 14 & -0.32 & 0.04 & 0.05 & 11 &   ...\tabularnewline
HD16815  &  ...  & -0.16 & 0.07 & 0.07 & -0.20 & 15 & -0.25 & 0.04 & 0.06 & 12 &   ...\tabularnewline
... & ... & ... & ... & ... & ... & ... & ... & ... & ... & ... & ...\tabularnewline
\hline 
\end{tabular}
\end{center}
\noindent

Notes: $^{*}$ The [X/H] abundances after correction for the $T_{eff}$ trends.

\end{table*}

\begin{figure*}
\begin{center}
\begin{tabular}{c}
\includegraphics[angle=0,width=0.8\linewidth]{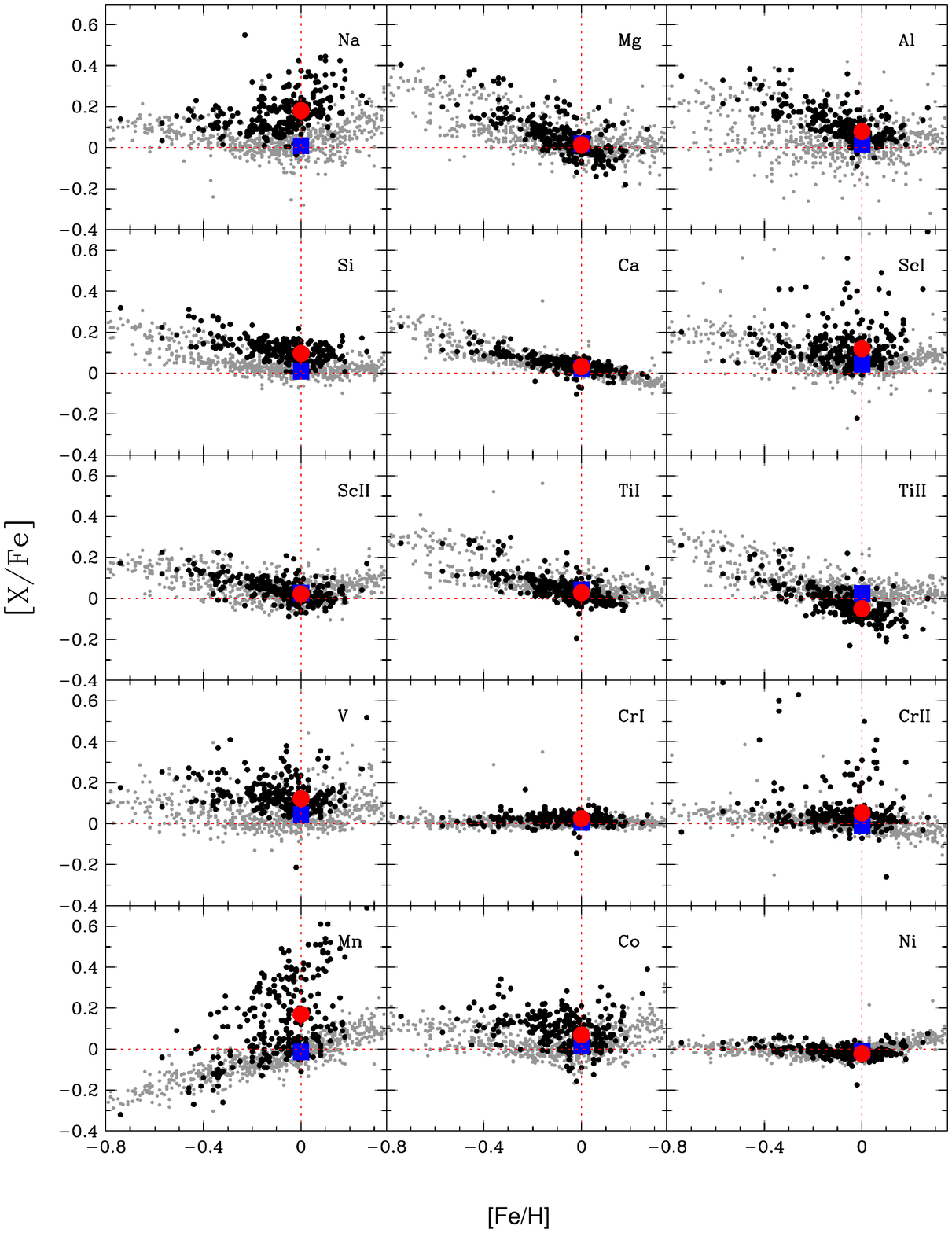}
\end{tabular}
\end{center}
\vspace{-1.3cm}
\caption{[X/Fe] vs. [Fe/H] plots. 
The black dots represent the stars of the sample and the gray small dots represent stars from \citet{Adibekyan-12b} 
with \emph{$T{}_{\mathrm{eff}}$} = \emph{T$_{\odot}$$\pm$$500$ K}. The red circle and blue square show the average [X/Fe] value 
of stars with [Fe/H] = 0.0$\pm$0.1 dex. 
Each element is identified in the \emph{upper right corner} of the respective plot.}
\label{fig_xfe_feh}
\end{figure*}

\subsection{Testing the validity of the stellar parameters}

Although \citet{Alves-15} have shown that the stellar parameters in general agree very well with the literature data, the consistency does
not always imply correctness. Moreover the stellar parameters were derived by completing an LTE abundance analysis and by using only iron lines.
To check the validity limit of the adopted methodology in terms of stellar parameter ranges
we tested our results in two ways \citep[see also][]{Adibekyan-12b}. 

First, in Fig.~\ref{fig_x1x2_param} we plot the [CrI/CrII], [ScI/ScII], and [TiI/TiII] as a function of the stellar parameters to ensure that the 
ionization equilibrium enforced on the FeII lines \citep{Alves-15} is acceptable to other elements. For comparison, the field FGK dwarf
stars from \citet{Adibekyan-12b} are presented. Most of the trends are nearly flat around zero in contradiction to their 
unevolved counterparts for which a gradual increase with decreasing \emph{$T{}_{\mathrm{eff}}$} was observed. For our stars only an increase of [ScI/ScII] ratio can be seen
with the decrease of \emph{$T{}_{\mathrm{eff}}$}. However, the results obtained for ScI and CrII should be considered with caution since their abundances have been derived 
by using only one line. From the figure one can notice a small offset from zero for [TiI/TiII] ratio and [ScI/ScII]. These positive offsets 
probably do not have relation to the NLTE effects as discussed in \citet{Bergemann-11} for [TiI/TiII] and still need to be understood.

\begin{figure*}
\begin{center}
\begin{tabular}{c}
\includegraphics[angle=0,width=0.6\linewidth]{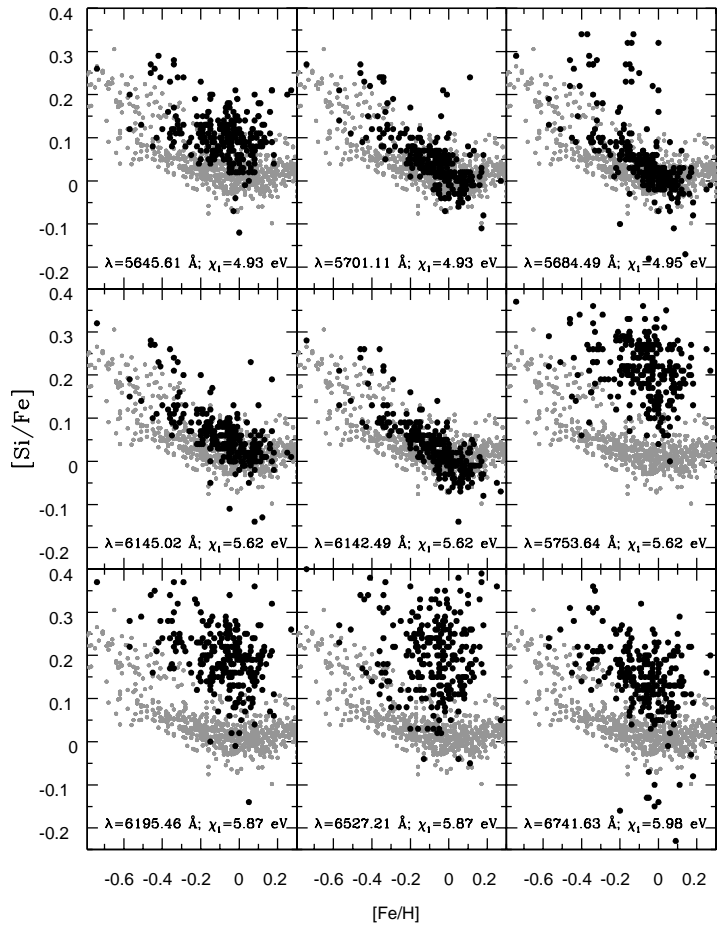}
\end{tabular}
\end{center}
\vspace{-0.8cm}
\caption{[Si/Fe] vs. [Fe/H] plots for Si lines of different excitation potentials. 
The black dots represent the stars of the sample and the gray small dots represent stars from \citet{Adibekyan-12b} 
with \emph{$T{}_{\mathrm{eff}}$} = \emph{T$_{\odot}$$\pm$$500$ K}. 
The wavelength of each line and excitation energy of the lower energy level ($\chi_{\mathrm{1}}$) is identified in the \emph{lower left corner} of the respective plot.}
\label{fig_si_fe_3x3}
\end{figure*}

\citet{Tsantaki-13} showed that by correcting stellar parameters (mainly \emph{$T{}_{\mathrm{eff}}$}), using carefully selected line-list especially designed for cool stars,
the observed trends of [XI/XII] with stellar parameters get flatter. For example, an overestimation of \emph{$T{}_{\mathrm{eff}}$} for cool stars might cause of the trends.
The dependence of the abundances of ionized and neutral species on the surface gravity is also discussed in \citet{Mortier-13a}. 
  
Another way of testing the stellar parameters is to plot [X/Fe] against \emph{$T{}_{\mathrm{eff}}$} (Fig.~\ref{fig_xfe_teff}). For the comparison the dwarf sample
of \citet{Adibekyan-12b} is also presented. Stellar evolutionary models do not suggest significant trends of these ratios with \emph{$T{}_{\mathrm{eff}}$}. However, for several elements 
we detected significant trends. To evaluate the significance of the trends we performed a linear fit and followed the procedure described in 
\citet{Figueira-13}. In short, first, we obtained the zero-centered distribution of the correlation coefficient by randomly bootstrapping (building random
samples by shuffling the parameters among the observed set of parameters) the observed data pairs 10$^{4}$ times. 
Then we calculated the correlation coefficient for each of these uncorrelated data sets and then the average and standard deviation of these values. By assuming a
Gaussian distribution for \textit{R} (correlation coefficient), we calculated the probability that the \textit{R} of the original
dataset was obtained by pure chance.
The significance of the trends and the slopes are presented in Table~\ref{table-2}. From the figure, one 
can see that for most cases the trends are less steep compared to those observed for the dwarfs%
\footnote{Note that the stellar parameters for the dwarfs were not derived by using the \citet{Tsantaki-13} line-list which
is especially designed for cool stars.}, which in turn speaks about the correctness of 
the stellar parameters used to derive the abundances.

\citet{Adibekyan-12b} already discussed several possible reasons for the observed trends of [XI/XII] with stellar parameters, and [X/Fe] with \emph{$T{}_{\mathrm{eff}}$}%
\footnote{Note that the \citet{Adibekyan-12b} sample essentially consists of dwarfs.} and
concluded that the observed trends are probably not an effect of stellar evolution, and uncertainties in atmospheric models are the dominant effect 
in the measurements. The authors afterwards  removed the \emph{$T{}_{\mathrm{eff}}$} trend as it was done also in other works \citep[e.g.,][]{Valenti-05, Petigura-11}.

\begin{figure*}
\begin{center}$
\begin{array}{ll}
\includegraphics[width=0.45\linewidth]{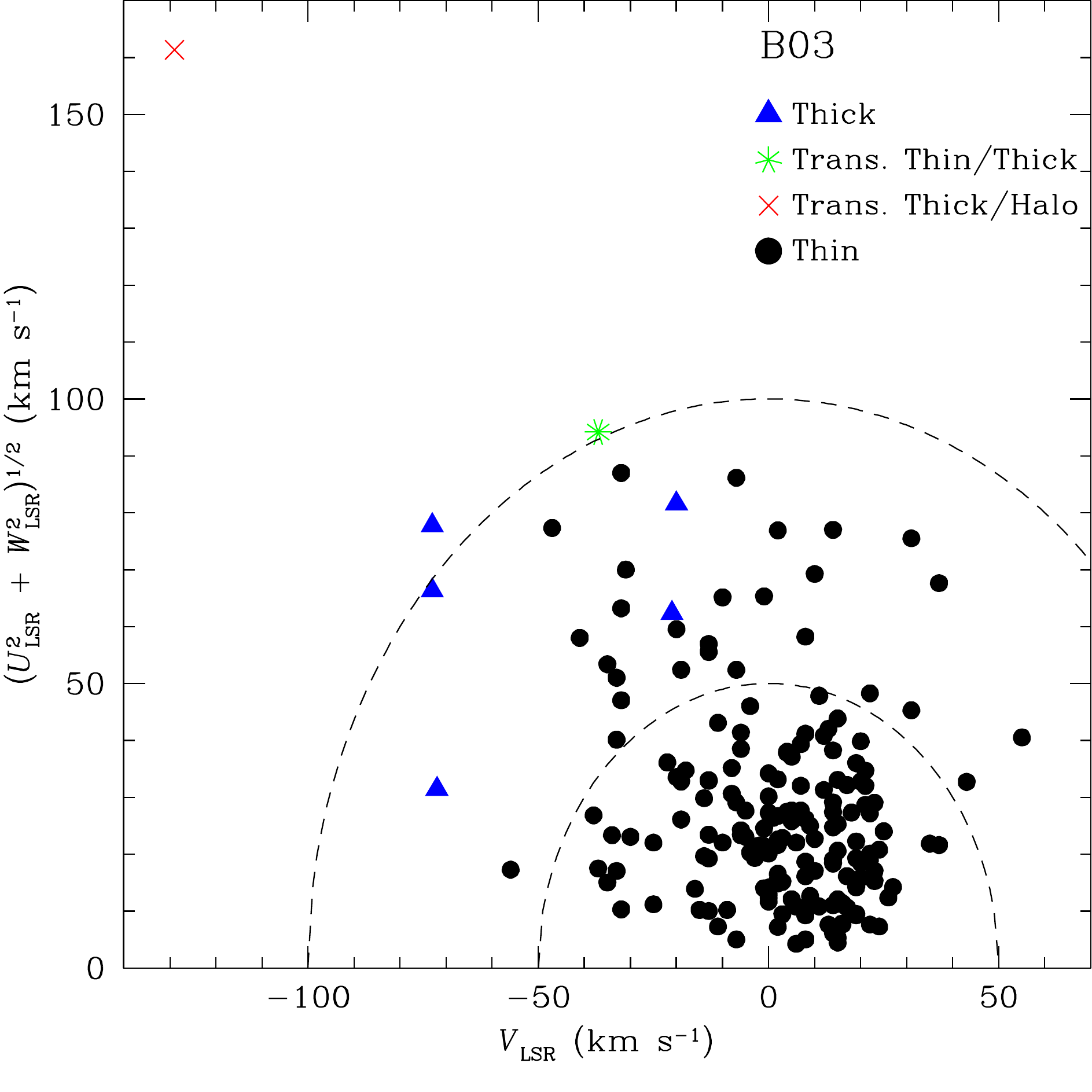}&
\includegraphics[width=0.45\linewidth]{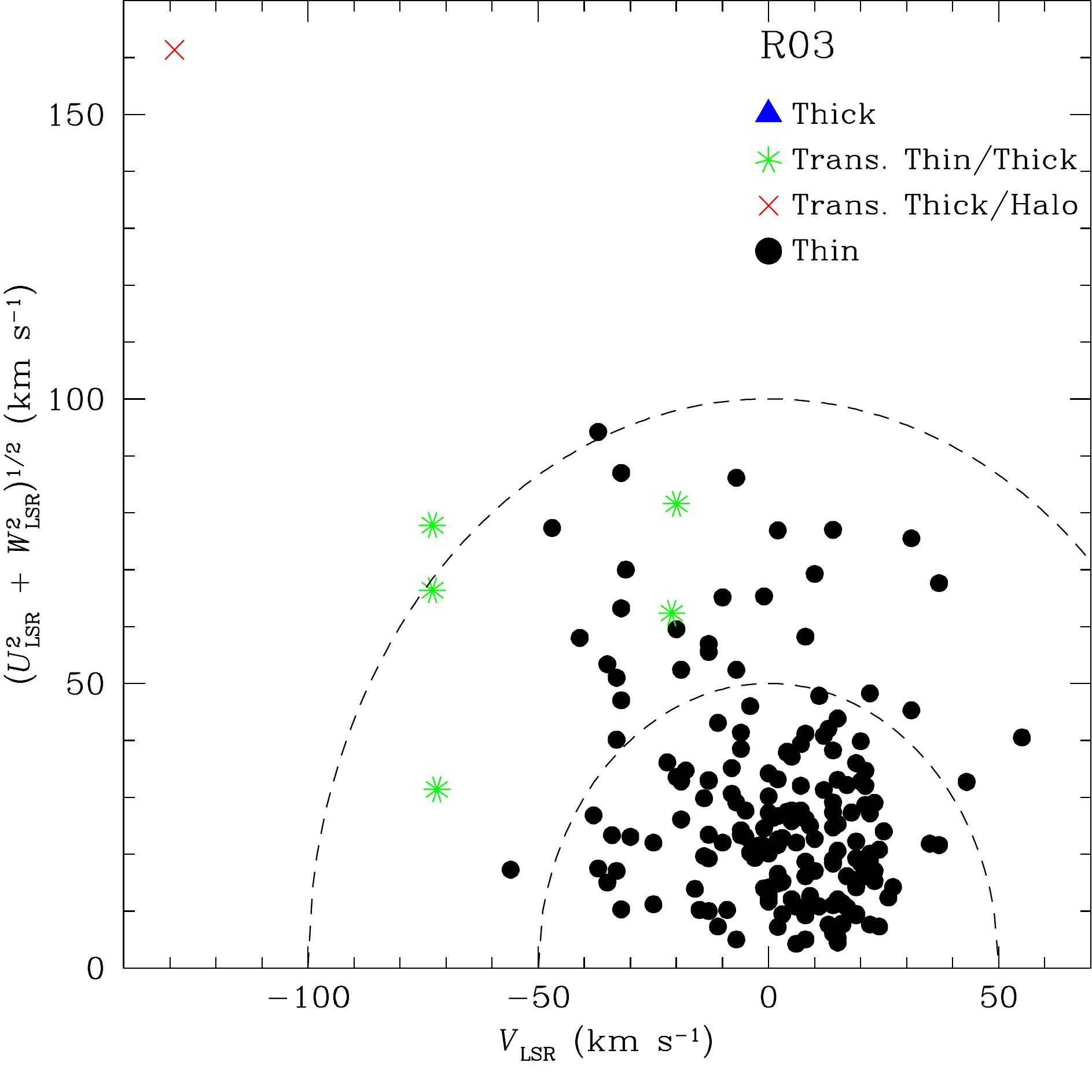}
\end{array}$
\end{center}
\vspace{-0.6cm}
\caption{Toomre diagram for the entire sample. The \textit{left} and \textit{right} panels show the separation of the stellar groups according to the
\citet[][B03]{Bensby-03} and \citet[][R03]{Robin-03} prescription, respectively. The symbols are explained in the figure. }
\label{fig_toomre}
\end{figure*}

Since by fitting the data and simply subtracting the fit would force the mean [X/Fe] to zero (which is an nonphysical situation), \citet{Adibekyan-12b}
added a constant term chosen so that the correction is zero at solar temperature. In our case the stars are cooler and their temperatures do
not reach the solar temperature so we decided to apply another approach. In this case, the constant term was chosen so that
the correction is zero at \emph{$T{}_{\mathrm{eff}}$} = 4960 K, which is the mean temperature of our sample stars. However, we appreciate the fact that this
approach and the choice  of the constant term is arbitrary. For this reason we decide to use the original (before detrending) chemical
abundances for the rest of our study. 

The dependence of [X/Fe] on stellar gravity and microturbulence and metallicity is shown in Fig.~\ref{fig_xfe_logg}, Fig.~\ref{fig_xfe_vtur},
and Fig.~\ref{fig_xfe_feh}, respectively. For most of the species we did not observe a trend with \emph{$\xi{}_{\mathrm{t}}$} and {$\log g$}, and some of 
the observed trends  probably reflect the correlation of \emph{$T{}_{\mathrm{eff}}$} with other stellar parameters (see Fig.~\ref{fig_param-param}).

As a final check, we compare our derived abundances with those obtained by \citet{Liu-07}. We note that this 
is the only literature source where we find enough stars (14 stars) in common to compare. We found very good 
agreement for all the species except vanadium: $\Delta [Na/H] = 0.02\pm0.12$, $\Delta [Mg/H] = -0.06\pm0.12$, $\Delta [Al/H] = 0.05\pm0.04$,
$\Delta [Ca/H] = 0.03\pm0.06$, $\Delta [Si/H] = 0.03\pm0.03$, $\Delta [Ti/H] = 0.04\pm0.08$, $\Delta [Ni/H] = -0.01\pm0.05$, and 
$\Delta [V/H] = 0.19\pm0.05$ %
\footnote{$\Delta [X/H] = [X/H]_{our} - [X/H]_{theirs}$.}.

The [X/H] abundances of all the stars before and after correction (if the significance of the correlation is above 3$\sigma$)
for the \emph{$T{}_{\mathrm{eff}}$} trends are available at the CDS (see also Table~\ref{table-abundance}).

\subsection{[X/Fe] vs. [Fe/H] relation. Galactic chemical evolution}

The [X/Fe] vs. [Fe/H] relation plot is traditionally used to study the Galactic chemical evolution because iron is a good chronological
indicator of nucleosynthesis. In Fig.~\ref{fig_xfe_feh} we present the dependence of [X/Fe] on metallicity for our sample of giant stars 
and for FGK dwarf stars from \citet{Adibekyan-12b}%
\footnote{Only stars with \emph{$T{}_{\mathrm{eff}}$} = \emph{T$_{\odot}$$\pm$$500$ K} are presented, because of the highest
accuracy in the parameters and chemical abundances in these stars.}. In the figure we also showed the average value of [X/Fe] for 
stars in the metallicity range of 0.0 $\pm$ 0.1 dex, where the Galactic chemical evolution effects are small.
As one can see, for all the elements the general behavior of [X/Fe] with the metallicity is similar for giant and dwarf stars, and  
 reflects the Galactic chemical evolution in the solar neighborhood. However, one can also clearly notice that, for some elements 
(Co, Na, V, Mn, Al, and Si) while the Galactic chemical evolution trends are similar, they are shifted:  for giant stars
having higher [X/Fe] values at a fixed metallicity. The largest offset is seen for Na and Mn, and a bit less in Si and Al.
In general, Na and Al are not good tracers of chemical evolution and affected by internal mixing processes in the stars.
The Mn abundance was obtained by using only one line and it should be considered with caution. Moreover, the scatter in [Mn/Fe] is very high,
indicating unrealistic abundances for some fraction of the stars.

Overabundances of sodium and aluminum (also silicon in some cases) in open cluster giants (compared to the abundances of dwarfs)
were already observed by several authors \citep[e.g.,][]{
Friel-03, Friel-05, Tautvaisiene-05, Jacobson-07, Villanova-09, Santrich-13}. In most of these studies, the trends were explained as a 
stellar evolutionary effect, due to the deep mixing produced by the hydrogen burning cycle, after stars have left the main sequence. 
For a complete picture, one should perform thorough analysis taking into account the non-LTE effects which are stronger
for giants stars and also the systematic errors which might arise due to particular spectroscopic analysis method used. 
For example, it is well known that sodium lines suffer from non-LTE effects which lead to an overestimation of the Na abundances 
\citep[e.g.,][]{Alexeeva-14}. In our analysis we used two sodium lines (at 6154.23 \AA and 6160.75{\AA}) which were studied
for non-LTE effects in \citet{Alexeeva-14}. The average EWs of these lines were $\sim$ 70 m{\AA} for 6154.23 {\AA}, and 
$\sim$ 80 m{\AA} for the 6160.75{\AA}  line. According to Alexeeva et al., the non-LTE correction for our stars should be 
from -0.1 to -0.15 dex, which is close to the difference in [Na/Fe] between giants and dwarfs observed in this study.

The difference in Al abundances ([Al/Fe]) between giants and dwarfs obtained for solar-metallicity stars is not large (0.07 dex - about 1$\sigma$ scatter), but
seems to increase at lower metallicities.
However it should be noted that direct comparison of the abundance ratios at lower metallicities is not straightforward, 
since the Galacic chemical evolution effects and the relative fraction of thin and thick disk stars can be dominant. 
Several authors studied the non-LTE effects on the formation of Al lines \citep[e.g.][]{Baumueller-96,
Baumueller-97, Menzhevitski-12}. They showed that the non-LTE correction of the Al abundances, derived from the subordinate doublet $\lambda\lambda$ 6696.03, 6698.68 {\AA}, 
is very small at solar metallicities, does not depend strongly on the surface gravity and only significant at temperatures above 6500 K \citep[][]{Menzhevitski-12}.

The next element for which we obtained small, but systematic difference between giants and dwarfs is Si. The abundance of Si is not expected to
be affected be extra mixing processes in the stars. The few studies of the Si abundances taking into account the non-LTE deviations showed that
the effect is significant only at very low metallicities \citep[e.g.][]{Bergemann-13} and the non-LTE correction for Si of the Sun is about -0.05 dex \citep{Sukhorukov-12}.
Since in this study for the Al abundance derivation we used several lines with different excitation potentials (EP), which means different atmospheric layers of
formation and hence different sensitivities to non-LTE deviations, we decided to analyze [Si/Fe] against [Fe/H] for each individual line.
In Fig.~\ref{fig_si_fe_3x3} we plot [Si/Fe] against [Fe/H] for 9 Si lines with the lowest, intermediate and highest excitation energy of the 
lower energy level ($\chi_{\mathrm{1}}$). At the first glance it looks like the lines with the highest $\chi_{\mathrm{1}}$ show the highest 
deviations from the abundances derived for the FGK dwarfs. However, as can be seen in the middle panel of the plot
the three lines with exactly the same $\chi_{\mathrm{1}}$  show different behavior, for $\lambda$ 5753.64 {\AA} showing the largest difference.

{Fig.~\ref{fig_si_fe_3x3} shows that the picture is complex and probably several process (NLTE, unresolved blends) are acting and affecting the abundances
at the same time. To select the ``best'' lines i.e. lines which give  similar average abundances to that obtained for the dwarfs, for all the 
lines we calculated the average [Si/Fe] abundance ratio obtained for all the giants with solar-metallicity $\pm$ 0.1 dex and compared that with the average
[Si/Fe] obtained for dwarfs \citep{Adibekyan-12b} with meallicities in the range of 0.0 $\pm$ 0.1 dex. Then we used the $rms$ (scatter) of the [Si/Fe] 
calculated for the dwarfs%
\footnote{The scatter obtained for the dwarfs by averaging the abundance of many lines is more realistic than the scatter obtained from one spectral line for giant stars.}
to quantify the observed differences (in n$\times\sigma$). We found five (out of 15) Si lines which give [Si/Fe] abundance similar to that obtained for the dwarfs 
(less than 2$\sigma$ difference).
The [Si/Fe] abundance obtained by averaging the abundances of the mentioned five lines against [Fe/H] is presented in the Fig.~\ref{fig_xfe_feh_best_lines}.
We note that we do not claim that these selected lines are not affected by non-LTE effects of unresolved blends, but they provide abundances 
similar to that obtained for dwarfs, which probably means that they are more realistic.

We repeated the aforementioned procedure for all the lines for each element and calculated the difference in [X$_{line}$/Fe] between giants and dwarfs for each line.
This differences,in n$\times\sigma$, is presented in the last column of the line-list table (available at CDS). The [X/Fe] vs. [Fe/H] relation obtained by using only the ``best'' lines
(less than 2$\sigma$ difference) is shown in Fig.~\ref{fig_xfe_feh_best_lines}. The corresponding [X/H] abundances are available at CDS.
 
\section{Kinematics and stellar populations}

It is  becoming increasingly clear  that a separation of  the Galactic stellar components  based  only on  stellar  abundances  
is  superior to  kinematic separation     \citep[e.g.,,][]{Navarro-11,     Lee-11,    Adibekyan-11,
Liu-12, Recio-Blanco-14}, because chemistry is a relatively more stable property of a star than its spatial positions
and kinematics. However, as mentioned above, some changes in abundances of some elements are expected when the stars
are evolving and leaving the main sequence. In this  analysis, to  separate the  thin-  and thick disk stellar components, we used the position of the 
stars in the [$\alpha$/Fe] - [Fe/H] plane (here $\alpha$ refers to the average abundance of Mg, Si, and Ti), 
but separately also a kinematics approach is applied. 

The space velocity components for 183 stars out of 257 were derived with respect to the local standard of rest (LSR), 
adopting the standard solar motion (U$_{\odot}$, V$_{\odot}$, W$_{\odot}$) = (11.1, 12.24, 7.25) km $\mathrm{s}^{-1}$ of 
\citet{Schonrich-10}. For the remaining 73 stars we did not calculate the velocities because of the deficit of astrometric literature data. 
The radial velocities,  parallaxes and proper motions were taken from the SIMBAD Astronomical Database%
\footnote{http://simbad.u-strasbg.fr/simbad/}.  Combining the measurement errors in the parallaxes,
proper motions, and radial velocities, the resulting average errors in the U, V, and W velocities are of about 2-3 km {s}$^{-1}$.

To assess the likelihood of the stars being a member of different Galactic populations, we followed \citet{Reddy-06}.
The probabilities that the stars belong to different stellar populations were calculated, having
adopted both the \citet{Bensby-03} and \citet{Robin-03} population fractions. 
We refer the reader to \citet{Adibekyan-12b} for the details of the computation.
The Galactic space velocity components and the probabilities to assign the stellar population to 
which the stars belong are available at the CDS.

According to the \citet{Bensby-03} criteria, among the 183 stars, we have 176 (96\%) stars from the thin disk, 5 from the thick
disk, and 2 stars are considered to be transition stars that do not belong to any group. Adopting the
criteria from \citet{Robin-03} gives 177 (97\%) thin disk stars, 5 star with kinematics suggesting a thick/thin disk transition, and 
one star with a classification of thick-disk/halo transition object. The distribution of the stars in the Tommre diagram is shown in 
Fig.~\ref{fig_toomre}.

As mentioned above, in addition to the difference in their kinematics, the thin- and thick disk stars are also different 
in their $\alpha$ content at a given metallicity \citep[e.g.,][]{Fuhrmann-98,Fuhrmann-08}. This dichotomy in the chemical evolution allows one to separate different stellar 
populations.

The [$\alpha$/Fe] versus [Fe/H] plot for the sample stars  along with the dwarf stars 
from \citet{Adibekyan-12b} with \emph{$T{}_{\mathrm{eff}}$} = \emph{T$_{\odot}$$\pm$$500$ K} is depicted in Fig.~\ref{fig_alfe_fe}%
\footnote{The chemical dissection of the disks is presented in the appendix.}. 
As one can see from the figure the two samples show similar trends, with giant stars having on average higher [$\alpha$/Fe] values 
at a fixed (low) [Fe/H]. This observed difference might arise from our assumption of LTE line formation. 
The non-LTE effects are stronger for metal-poor stars, but these effects depends also on other stellar parameters (e.g. gravity and temperature)
and also they different for different elements and they differ from line to line. Thus, for fully understanding of the main reason of the observed 
abundance difference between giants and dwarfs a complete non-LTE analysis is needed.

Our chemical separation of the Galactic disks suggest that 23 stars (9\%) in the sample show enhanced $\alpha$ abundances.
In \citet{Adibekyan-11} and \citet{Adibekyan-13} the high-$\alpha$ stars were separated into two groups with a gap in both 
[$\alpha$/Fe] and metallicity. It is interesting to see that the gap in [Fe/H] for high-$\alpha$ stars can be also seen
in our sample at the same metallicity ($\approx$ -0.2 / -0.3 dex). Following the same logic and definitions as in \citet{Adibekyan-13},
the 10 stars with enhanced [$\alpha$/Fe] and [Fe/H] below -0.3 dex can be classified as thick-disk stars, and the remaining 13
stars as high-$\alpha$ metal-rich stars (h$\alpha$mr). With this definition we see that 4\% of the stars belong to the Galactic thick disk, as the 
kinematic separation was suggesting.

We note that the current sample is small and we will avoid of a definite conclusion about the
existence of the mentioned ``gap`` at [Fe/H] $\approx$ -0.3 dex and the distinction of the two $\alpha$-enhanced metal-poor and metal-rich populations. 
However, the fact that the two different homogeneously analyzed samples (the current one and the one from \citet{Adibekyan-11}) show 
quite similar features probably is more than just a hint about the existence of the h$\alpha$mr stars as a distinct stellar family.
Moreover, recent study of an inner disk metal-rich open cluster, Berkeley 81, shows that the stars are enhanced in $\alpha$ element
Magrini et al. (2014, in prep.), thus confirming that the h$\alpha$mr stars have inner disk origin as suggested by \citet{Adibekyan-13}.
However, we want to note that no similar ''gap`` was found in \citet{Bensby-14} where the authors suggested that the h$\alpha$mr stars present 
the metal-rich tail of the thick disk. As mentioned in \citet{Bensby-14}, a large sample with well-controlled selection function
(e.g., Gaia-ESO survey - \citet{Gilmore-12}) would help us to understand the real nature of the h$\alpha$mr stars.

\begin{figure}
\begin{center}
\begin{tabular}{c}
\includegraphics[angle=270,width=1\linewidth]{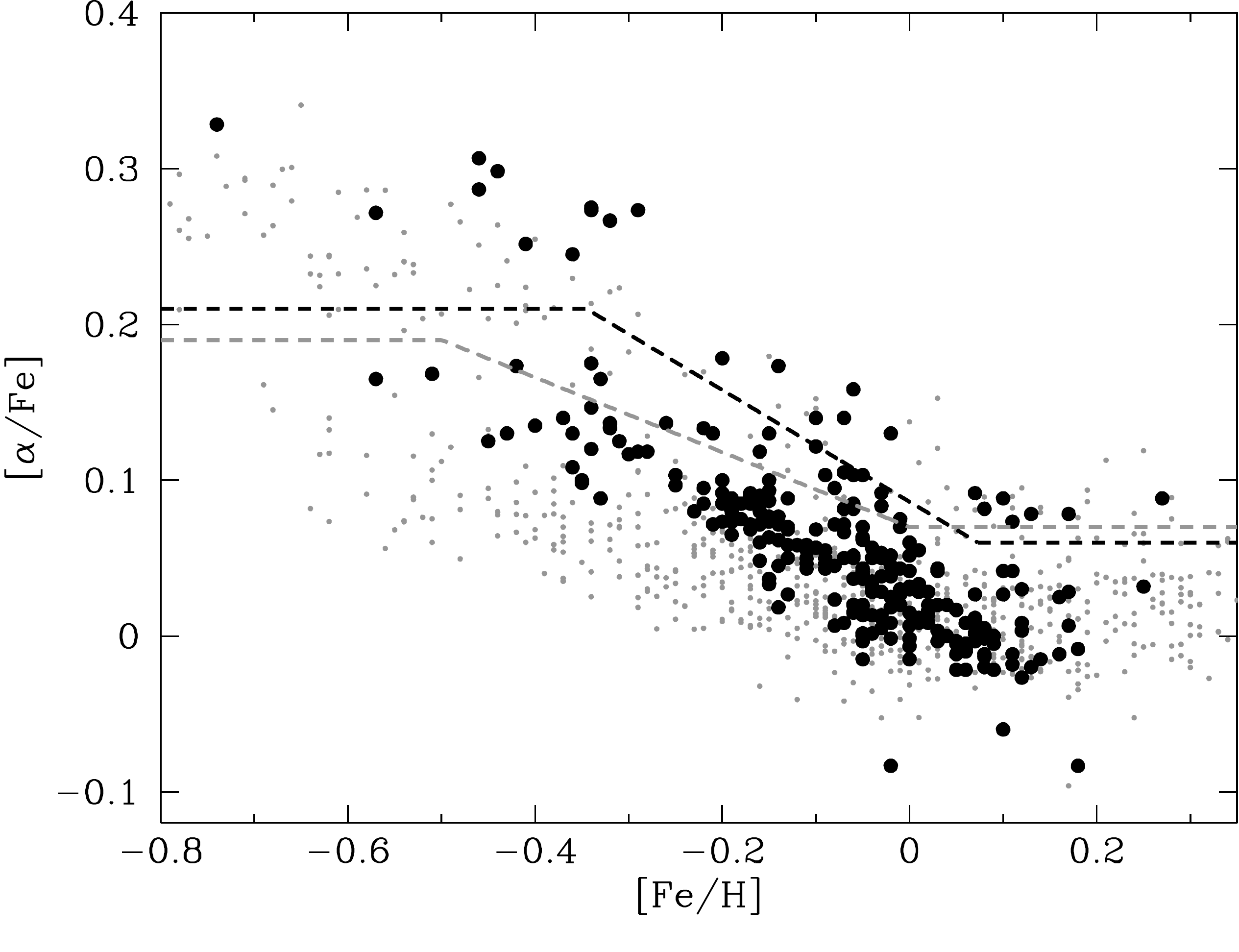}
\end{tabular}
\end{center}
\vspace{-0.6cm}
\caption{[$\alpha$/Fe] vs. [Fe/H] for the current sample (black dots) and for the stars from \citet{Adibekyan-12b} 
with \emph{$T{}_{\mathrm{eff}}$} = \emph{T$_{\odot}$$\pm$$500$ K} (gray small dots). The separation between the thick- and thin-disk
stars for the two samples are presented in black and gray dashed lines.} 
\label{fig_alfe_fe}
\end{figure}

\section{Metallicity distribution}

As mentioned above, several authors tried to understand the reason why the apparent giant-planet -- metallicity correlation does not
exist for evolved stars. As recently suggested by \citet{Mortier-13}, a possible reason might be a selection bias due to 
{\it B -- V} colour cut-off. 

In Fig.~\ref{fig_logg_feh}, we plotted the relation between stellar metallicity and surface gravity.
For the comparison, the dwarf stars sample from \citet{Adibekyan-12b} is also presented. From the figure one can easily see that
the giant stars sample lacks high-metallicity and low-gravity stars, and also low-metallicity and high-gravity stars. This is 
again probably because of the selection criteria used to define the sample.

To avoid the issues related to the selection effects, an unbiased giant sample with no colour cut-off and homogeneously
derived parameters is needed that is systematically searched for planetary companions. However, it is still possible to
overcome the effect of the {\it B -- V} colour cut-off if one considers, for example, only stars in the ''cut rectangle`` shown in 
Fig.~\ref{fig_logg_feh} (red rectangle), where the stars are equally distributed. However, these ''cut rectangles`` will consist of 
stars with narrower ranges of metallicities (from -0.25 to 0.15 dex in the example of Fig.~\ref{fig_logg_feh}), which is also
an issue since the giant-planet -- metallicity correlation is more pronounced at high metallicities (at least for dwarf stars). 

In the right panel of Fig.~\ref{fig_logg_feh}, we show the metallicity distribution of giant and dwarf stars where  narrower [Fe/H]
distribution of giants is apparent. The figure also shows that the two distributions are peaked at similar metallicities, close to the solar value.
The median (and its standard deviation) of the metallicity distributions of giant and dwarf stars are -0.05 (0.18) dex and -0.10 (0.33) dex, respectively%
\footnote{We note that the standard deviation of the median is calculated as 1.25*$\sigma$, where $\sigma$ is the standard deviation of the distribution.}. 
Several studies have already observed this tendency of evolved stars lacking the metal-rich and
very metal-poor tails \citep[e.g.,][]{Taylor-05, Takeda-08, Luck-07, Ghezzi-10}. 

The stars in this sample have stellar masses between 1.5 and 4.0 M$_{\odot}$ \citep{Alves-15}, and hence should be on average younger
than the dwarfs from \citet{Adibekyan-12b}. The younger age together with the age -- metallicity dispersion relation 
\citep[e.g.,][]{daSilva-06, Haywood-08, Casagrande-11, Maldonado-13} might explain the narrower [Fe/H] distribution of the giants. Young stars are mostly 
local since they do not have time to migrate within the Galaxy \citep[][]{Wang-13, Minchev-13}. Radial migration in the disk
is expected to make the metallicity distribution wider, but does not change the mean abundance  \citep[][]{Wang-13}, as we see in Fig.~\ref{fig_logg_feh}. This is because 
mostly massive stars contribute to the chemical enrichment of the interstellar medium and they contribute mainly around 
their birth places because of their very short lifetime. The lack of very metal-rich giants can be understood along the same 
migration process, most of the old stars which migrate would come from the inner, metal-rich disk \citep[][]{Wang-13, Minchev-13}.

In addition to the aforementioned astrophysical explanation, we would like to note again the selection effects which may
arise in evolved star samples due to {\it B -- V} colour cut-off. This selection bias may also make the metallicity distribution narrower.

\begin{figure}
\begin{center}
\begin{tabular}{c}
\includegraphics[angle=270,width=1\linewidth]{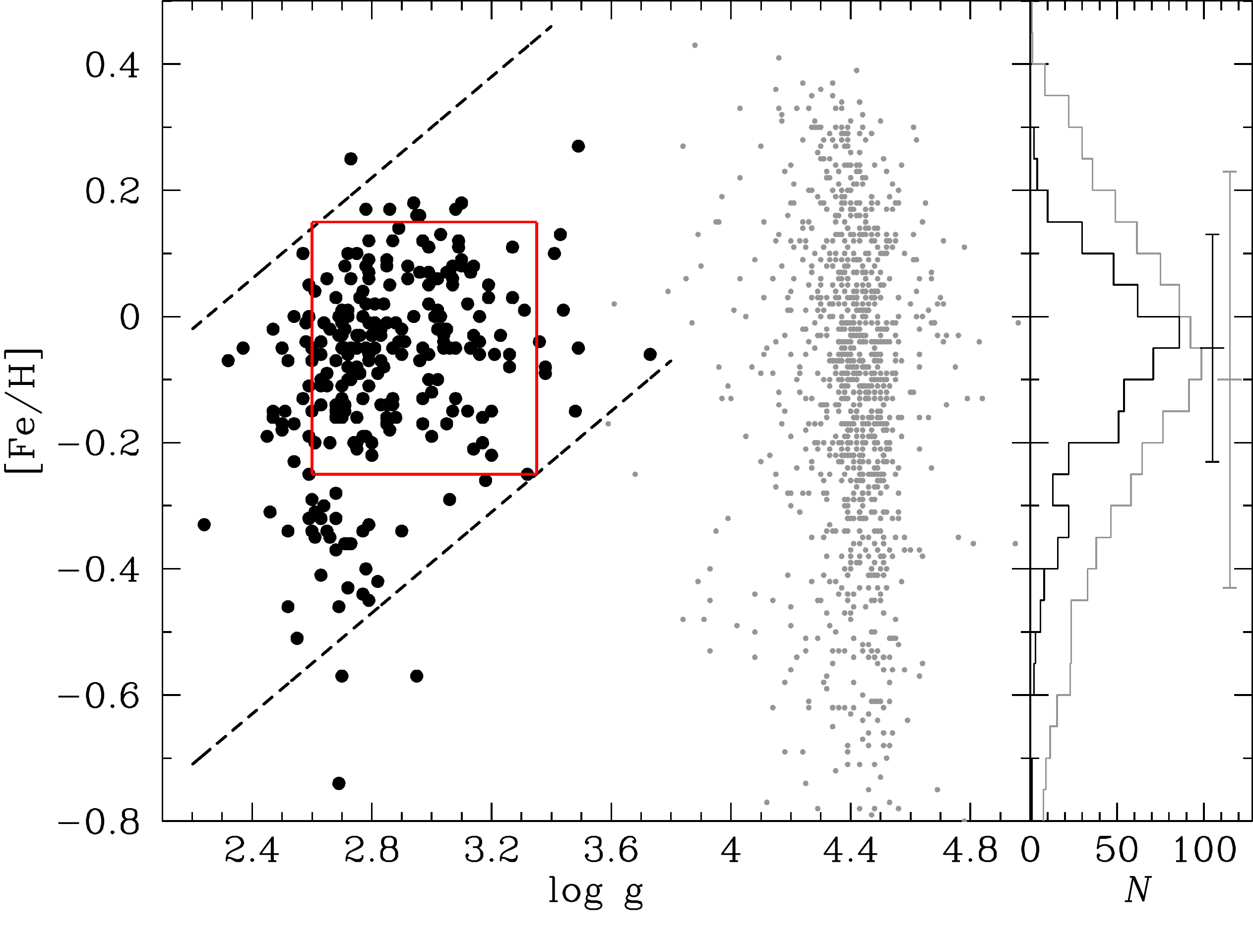}
\end{tabular}
\end{center}
\vspace{-0.6cm}
\caption{{\it Left panel}:[Fe/H] vs. $\log g$ for the current sample (black dots) and for the stars from \citet{Adibekyan-12b} 
(gray dots). The two black dashed lines were drawn by eye and show the biases
in the samples due to the B -- V cut-off. {\it Right panel}: The metallicity distribution of the two aforementioned samples.
The distribution of the giants stars (gray line) was multiplied by 2 for the better visual comparison. The median and its standard deviation is also presented
for metallicity distributions of both giants and dwarfs.}
\label{fig_logg_feh}
\end{figure}

\section{Summary of the results}

We have carried out a uniform abundance analysis for 12 refractory elements for a sample of 
257 field G-, K-type evolved stars that are being surveyed for planets using precise radial--velocity measurements with the 
CORALIE spectrograph. The abundances were derived using a carefully selected line-list and are based on the precise spectroscopic 
parameters derived by \citet{Alves-15} using the same spectra as were used in the present study.

We found that for all the elements Galactic chemical evolution trends are similar for giant and dwarf stars, while for some species
[X/Fe] values are shifted towards higher values at a fixed metallicity.
Our LTE analyis confirms the overabundance of Na in giant stars compared to the field FGK dwarf
stars from \citet{Adibekyan-12b}. This overabundance may have a stellar evolutionary character, even though the possible departures from
non-LTE may produce an enhancement of a similar degree \citep{Alexeeva-14}. We showed that an observed small overabundance of Si compared to the field FGK dwarf
vanishes when a shorter, carefully select line-list is used.

To separate Galactic stellar populations, we applied both a purely kinematical approach and chemical method. 
Our chemical separation suggests that 91\% of the stars, being $\alpha$-poor, belong to the thin disk and the 
remaining 9\% of the stars show enhanced $\alpha$-element abundances at a fixed [Fe/H]. This sample (while being not very large)
also suggests a ''gap`` in [Fe/H]  for high-$\alpha$ stars as observed in \citet{Adibekyan-11}. Following the definition of the last 
authors 4\% of the stars were classified as thick-disk members (being metal-poor) and 5\% as h$\alpha$mr stars.

The metallicity distribution of the giant stars is shown to be narrower than that of their non-evolved dwarf counterparts
\citep[see also][]{Taylor-05, Takeda-08}, but peaked at almost solar metallicity as in case of the dwarfs. The lack of very metal-rich
and metal-poor stars can be explained by the fact that most of the stars are originated in the solar vicinity.  
Evolved stellar samples mostly consist of massive stars, which have shorter lifetime than the dwarfs, and therefore do
not have enough time to migrate from further inner/outer disks \citep{Wang-13, Minchev-13}.

Our present sample, as most of the giant star samples searched for planets, is affected by {\it B -- V} colour cut-off which
excludes low-$\log$ g stars with high-[Fe/H] and high-$\log$ g stars with low metallicity. As discussed in \citet{Mortier-13},
this selection bias might be the reason of the absence of the correlation between occurrence of giant-planet planets and
stellar metallicity. We suggest to use stars in a ''cut-rectangle`` in the {$\log g$} -- [Fe/H] diagram to overcome the
aforementioned issue, if an unbiased sample is not available on hand.   

Although the current sample still contains only one star known to orbit a planetary companion \citep[HD 11977 --][]{Setiawan-05} most of the stars have already been
periodically observed over the last years. Before a significant number of planets are detected, this sample can be used as
a homogeneous comparison sample to study planet occurrence around giant stars. However, when exploring chemical
peculiarities of planet-hosting giant stars, one should bear in mind the chemical properties of these evolved stars discussed in
this paper (e.g., enhancement in Na, Al, etc.).

\section*{Acknowledgments}

This work was supported by the European Research Council/European Community under the FP7 through Starting Grant agreement 
number 239953. This work was also supported by the Gaia Research for European Astronomy Training (GREATITN) Marie Curie network, 
funded through the European Union Seventh Framework Programme ([FP7/2007-2013]) under grant agreement number 264895.
V.Zh.A. and S.G.S acknowledge the support from the Funda\c{c}\~ao para a Ci\^encia e a Tecnologia, FCT (Portugal) in the form of 
the fellowships SFRH/BPD/70574/2010 and SFRH/BPD/47611/2008, respectively. 
N.C.S was supported by FCT through the Investigator FCT contract reference IF/00169/2012 and POPH/FSE (EC) by FEDER funding 
through the program ``Programa Operacional de Factores de Competitividade'' - COMPETE. 
Research activities of the Obsevational Stellar Board of the Federal University of Rio Grande do Norte are supported
by continuous grants of CNPq and FAPERN Brazilian agencies and by the INCT-INEspaço.
S. A. acknowledges Post-Doctoral Fellowship from the CAPES brazilian agency (BEX-2077140), and also support by Iniciativa Cient\'ifica Milenio 
through grant IC120009, awarded to The Millennium Institute of Astrophysics.
G.I. acknowledges financial support from the Spanish Ministry project MINECO AYA2011-29060.
AM received funding from the European Union Seventh Framework Programme (FP7/2007-2013) under grant agreement number 313014 (ETAEARTH).
This research has made use of the SIMBAD database operated at CDS, Strasbourg, France.
We thank Mahmoudreza Oshagh for his interesting comments related to Fig. 6, and Elisa Delgado Mena for a very constructive discussion.
We would also like to thank the anonymous referee for useful comments that helped to improve the paper.

\bibliography{refbib}

\appendix

\section{Interdependence of stellar parameters and the microturbulence}

The interdependence of the fundamental parameters are presented in Fig.~\ref{fig_param-param}. 
The figure reveals several interesting correlations between the parameters, for instance one can see that the metallicity correlates with 
surface gravity (see also Sect. 5) and also stars with higher \emph{$T{}_{\mathrm{eff}}$} (above 5100 K)
show higher metallicity. Microturbulent velocity correlates with the {$\log g$}.
The significance of the observed correlations are estimated following the method described in \citet{Figueira-13} and \citet{Adibekyan-13} 
and the parameters of the linear relations are presented in  Table~\ref{table-1}. We note that five stars classified as ``outliers'' in the 
\emph{$\xi{}_{\mathrm{t}}$}-$\log g$, were excluded from the estimation of the significance of the correlations (see next section for details).

\begin{figure*}
\begin{center}
\begin{tabular}{c}
\includegraphics[angle=0,width=0.8\linewidth]{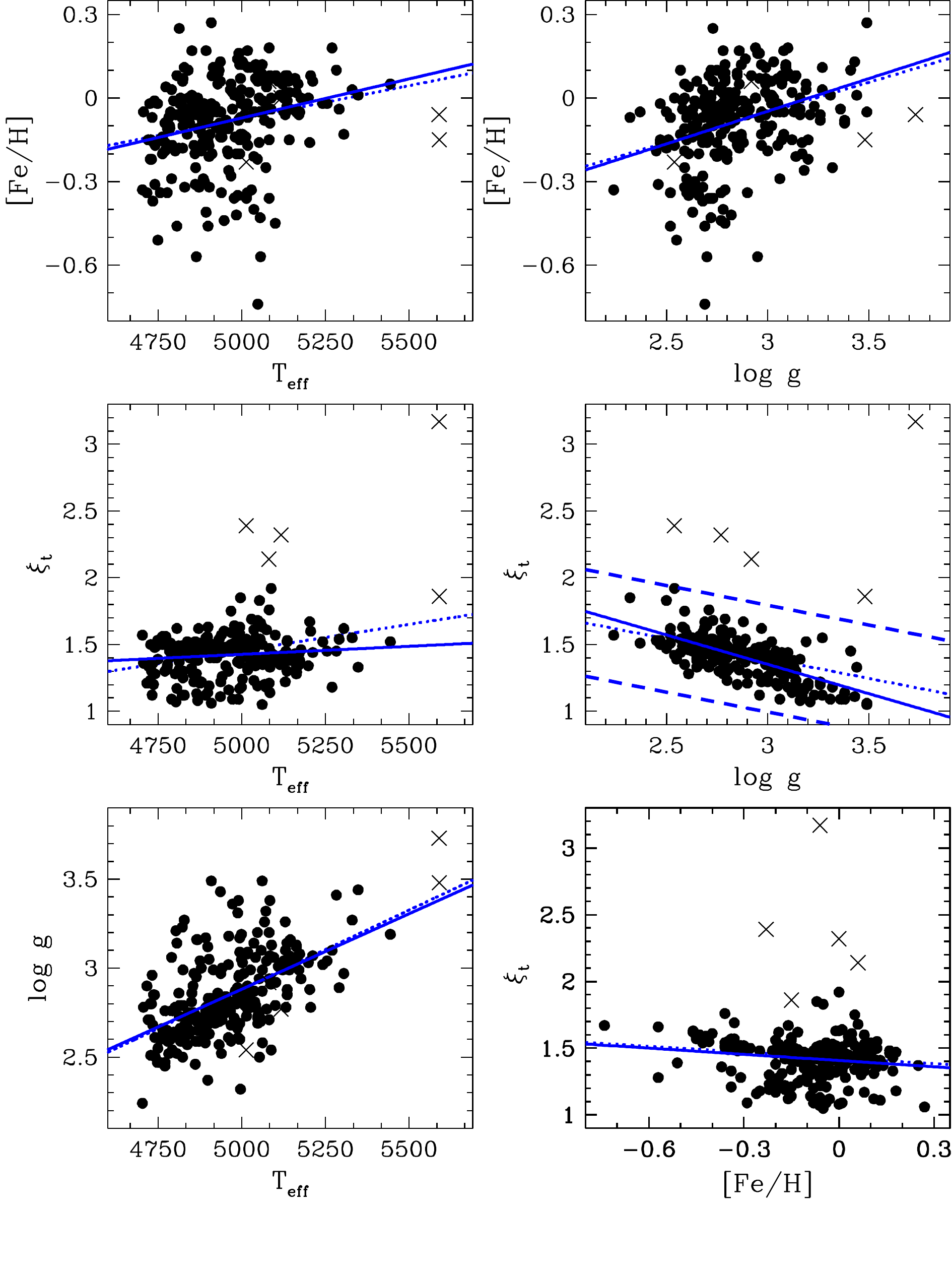}
\end{tabular}
\end{center}
\vspace{-0.8cm}
\caption{Interdependence of the stellar atmospheric parameters of the sample stars. The blue dotted lines depict the linear fits of the full data, and 
the sold lines are the fits after removing five ``outliers''. The 2$\sigma$ intervals of the linear fit of the \emph{$\xi{}_{\mathrm{t}}$}--$\log g$
relation are shown in blue dashed lines. The black croses indicate the five outliers.}
\label{fig_param-param}
\end{figure*}

\subsection{The microturbulence relationship}

Sometimes, when the number of iron lines is not large enough, a correct determination of microturbulence becomes very difficult because of small EW interval
of the FeI lines \citep[e.g.,][]{Mortier-13b}. In these cases, one uses empirically obtained relations between microturbulence and other stellar parameters. 
Several studies have shown that for FGK dwarf stars, microturbulent velocity depends on {$\log g$} and \emph{$T{}_{\mathrm{eff}}$} 
\citep[e.g.,][]{Nissen-81, Prieto-04, Adibekyan-12a, Tsantaki-13, Ramirez-13}. \citet{Takeda-08} has already suggested that the microturbulence correlates with the 
surface gravity, however the authors did not provide the analytic form of the relation. 

To find out the parameters the \emph{$\xi{}_{\mathrm{t}}$} correlates with, we first applied a linear fit for three pairs of data-sets: \emph{$\xi{}_{\mathrm{t}}$}-[Fe/H], 
\emph{$\xi{}_{\mathrm{t}}$}-$\log g$, \emph{$\xi{}_{\mathrm{t}}$}-\emph{$T{}_{\mathrm{eff}}$}. Then we evaluated the significance of the correlation, by using 
a bootstrap procedure as it was done
in \citet{Figueira-13}. As expected the strongest correlation is observed with $\log g$ (5.7$\sigma$), $\approx$4$\sigma$ in case of \emph{$T{}_{\mathrm{eff}}$},
and  $\approx$1.8$\sigma$ for [Fe/H]. However, the fits can be affected by the presence of several outliers as can be seen in Fig.~\ref{fig_param-param}. To
remove the outliers we used the \emph{$\xi{}_{\mathrm{t}}$}-$\log g$ relation (since it shows the strongest correlation), by applying 2$\sigma$-clipping 
(two times of residual standard deviation). 
Then, after cleaning the data from outliers we again fitted the data and again evaluated the significance of the relations. 
We found that microturbulence significantly correlated with the surface gravity (at about 11$\sigma$ level), and with the metallicity 
but with less degree of significance. The five ouliers were responsible for the ``strong'' relation observed between \emph{$\xi{}_{\mathrm{t}}$}
and \emph{$T{}_{\mathrm{eff}}$}.

After this test, we decided to present the relation of microturbulence only with $\log g$ and [Fe/H], which has the following functional form:

\begin{eqnarray}\nonumber
\xi_{t} & = & 2.72(\pm 0.08) - 0.457(\pm0.031)\times\log g \left. \right.\\
&& \left. + 0.072(\pm0.044) \times [Fe/H]  \,\  \right.
\end{eqnarray}

We note that this empirical relation is valid only for the range of stellar parameters that the stars in our sample cover.

\begin{table}
\centering
\caption{The coefficients of the linear fits (y = a $\times$ X +b) of the relations between the stellar parameters, along with the correlation coefficient
 and the significance. The number of stars is 251.}
\label{table-1}
\begin{tabular}{lcccc}
\hline 
\hline
\noalign{\vskip0.01\columnwidth} 
Elem & a & b & R$^{2}$  & z-score\tabularnewline
\hline 
\emph{$\xi{}_{\mathrm{t}}$} -- \emph{$T{}_{\mathrm{eff}}$} & 0.120$\pm$0.065 & 0.825$\pm$0.326 & 0.013  & 1.7 \tabularnewline
$\log g$ -- \emph{$T{}_{\mathrm{eff}}$} & 0.847$\pm$0.089 & -1.356$\pm$0.441 & 0.266 & 7.9 \tabularnewline
{[Fe/H]} -- \emph{$T{}_{\mathrm{eff}}$} & 0.280$\pm$0.069 & -1.472$\pm$0.343 & 0.061 & 3.9 \tabularnewline
{[Fe/H]} -- $\log g$ & 0.234$\pm$0.041 & -0.751$\pm$0.116 &  0.116 & 5.4 \tabularnewline
\emph{$\xi{}_{\mathrm{t}}$} -- $\log g$ & -0.440$\pm$ 0.029 & 2.673$\pm$0.083 & 0.476 & 10.8 \tabularnewline
\emph{$\xi{}_{\mathrm{t}}$} -- {[Fe/H]} & -0.154$\pm$0.057 & 1.407$\pm$ 0.010 & 0.027  & 2.6 \tabularnewline

\hline 
\end{tabular}
\end{table}

\section{[X/Fe] dependence on stellar parameters}

In this section we present [X/Fe] vs. $\log g$ (Fig.~\ref{fig_xfe_logg}), [X/Fe] vs. microturbulence (Fig.~\ref{fig_xfe_vtur}), 
and [X/Fe] vs. [Fe/H]  (Fig.~\ref{fig_xfe_feh_best_lines}) plots derived from the ``best'' lines as discussed in the main text.

\begin{figure*}
\begin{center}
\begin{tabular}{c}
\includegraphics[angle=0,width=1\linewidth]{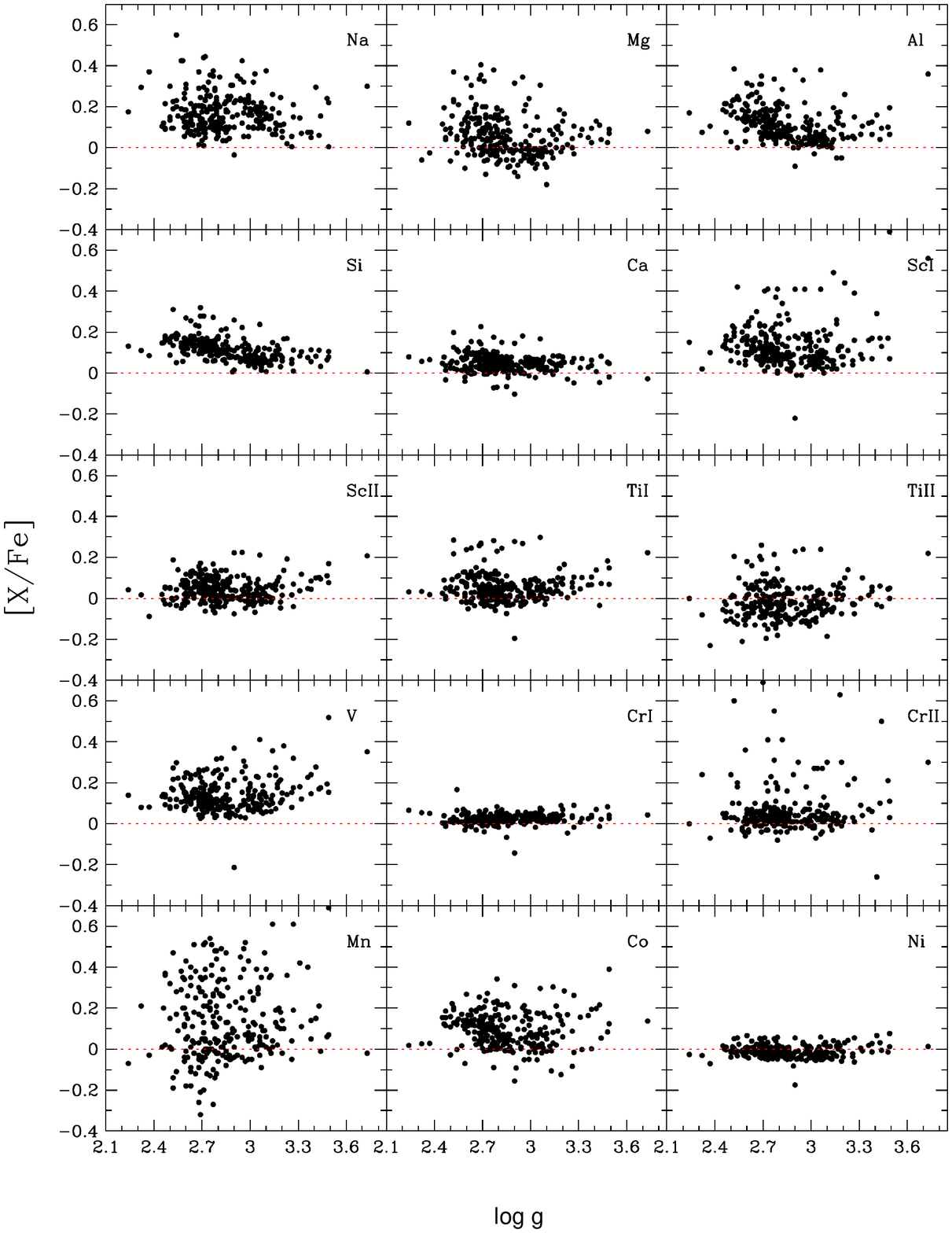}
\end{tabular}
\end{center}
\vspace{-0.6cm}
\caption{[X/Fe] vs. $\log g$  plots. Each element is identified in the \emph{upper right corner} of the respective plot.
The black dots represent the stars of the current sample. }
\label{fig_xfe_logg}
\end{figure*}

\begin{figure*}
\begin{center}
\begin{tabular}{c}
\includegraphics[angle=0,width=1\linewidth]{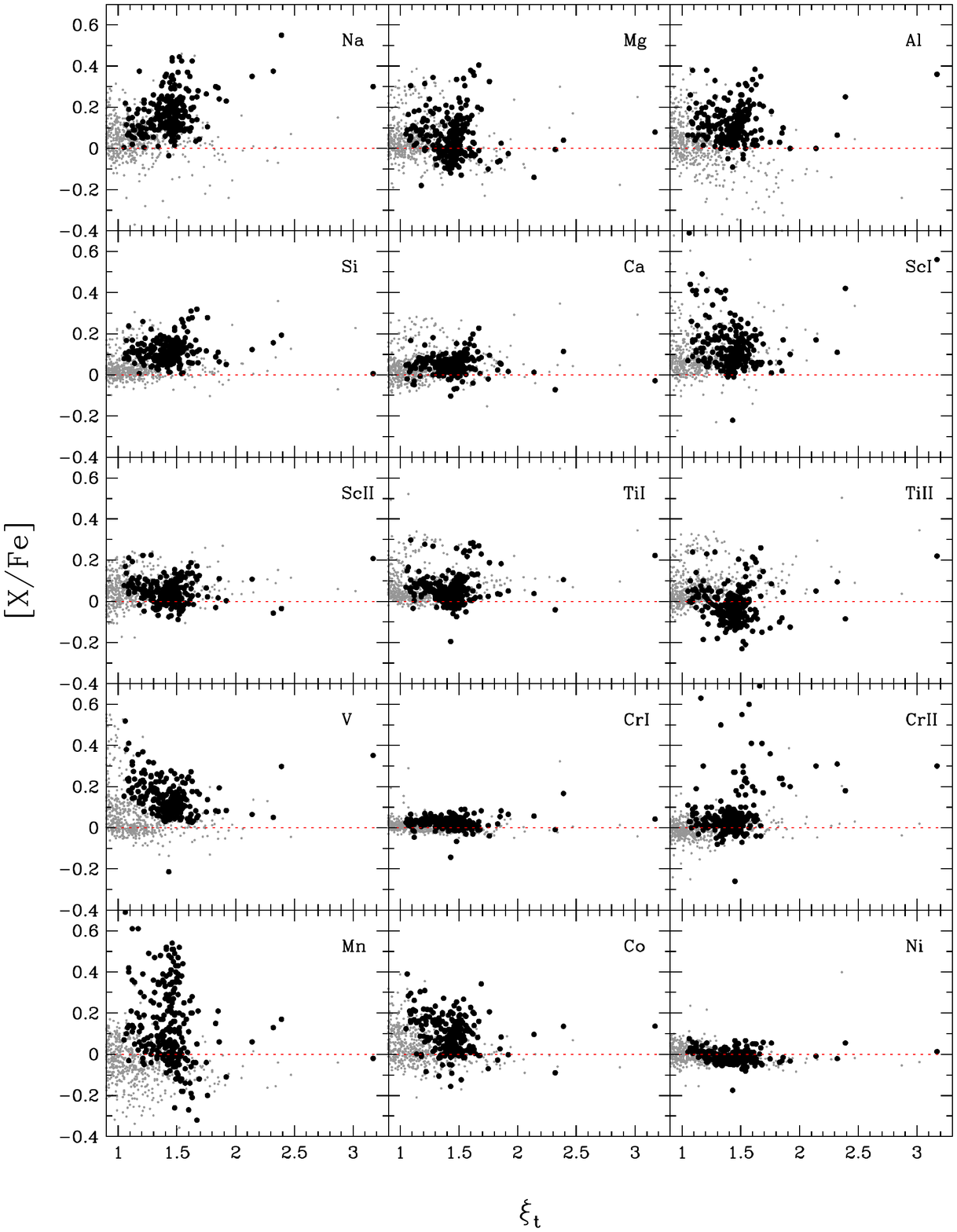}
\end{tabular}
\end{center}
\vspace{-1.3cm}
\caption{[X/Fe] vs. microturbulence  plots. Each element is identified in the \emph{upper right corner} of the respective plot.
The black dots represent the stars of the sample and the gray small dots represent stars from \citet{Adibekyan-12b}. }
\label{fig_xfe_vtur}
\end{figure*}

\begin{figure*}
\begin{center}
\begin{tabular}{c}
\includegraphics[angle=0,width=1\linewidth]{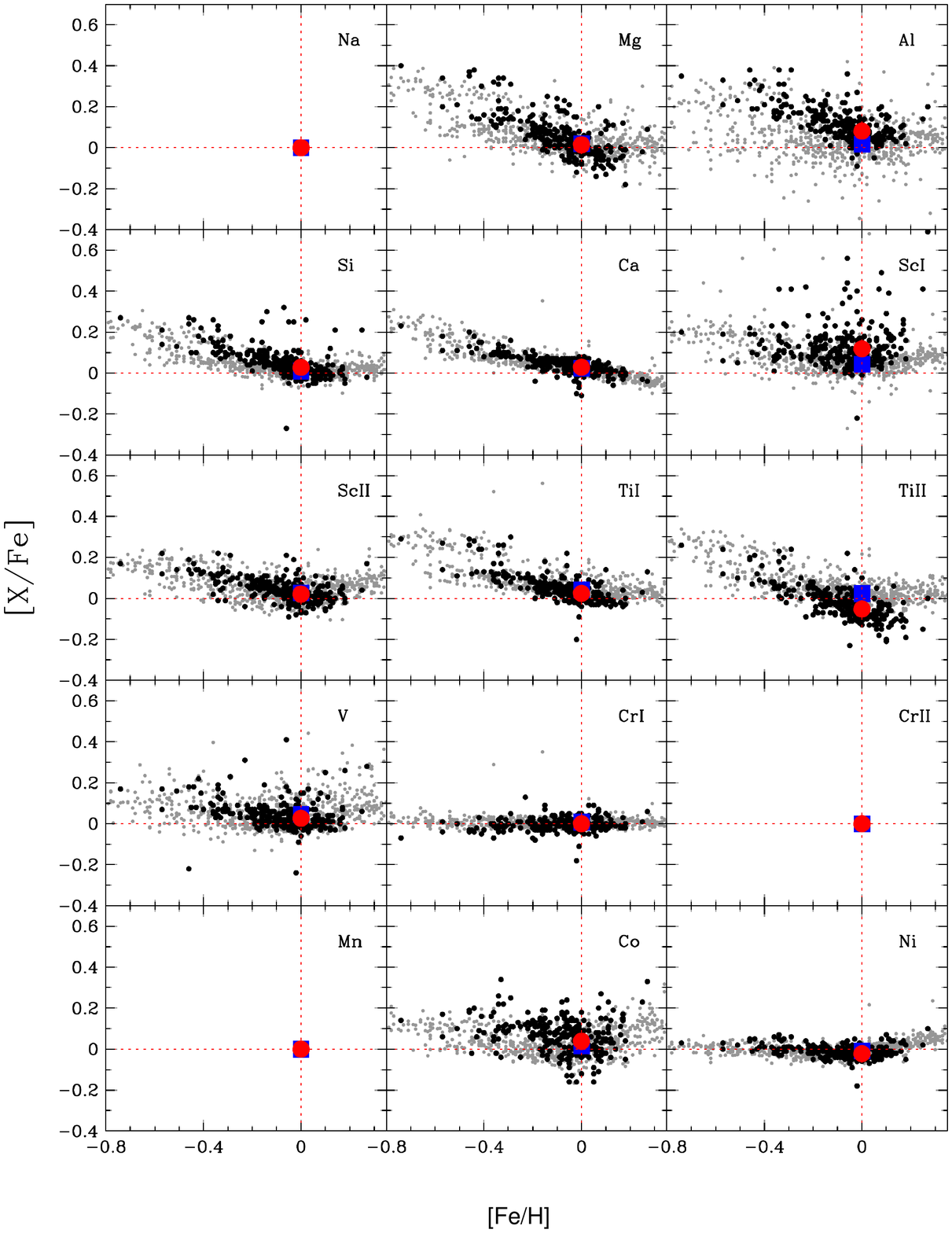}
\end{tabular}
\end{center}
\vspace{-1.3cm}
\caption{[X/Fe] vs. [Fe/H] plots derived from the ``best'' lines. For NaI, CrII, and MnI there was no ``best'' line(s) foud.
The black dots represent the stars of the sample and the gray small dots represent stars from \citet{Adibekyan-12b} 
with \emph{$T{}_{\mathrm{eff}}$} = \emph{T$_{\odot}$$\pm$$500$ K}. The red circle and blue square show the average [X/Fe] value 
of stars with [Fe/H] = 0.0$\pm$0.1 dex. 
Each element is identified in the \emph{upper right corner} of the respective plot.}
\label{fig_xfe_feh_best_lines}
\end{figure*}

\section{Separation of the Galactic disks by $\alpha$-enhancement}

For the separation of Galactic stellar population by the chemical properties of the stars was done following the method 
presented in \citet{Adibekyan-11}. We first divided the sample into three metallicity bins: [Fe/H] $<$ -0.3 dex, [Fe/H] $>$ 0.0 dex, and stars
in between. For the lowest and highest metallicity bins we easily identified the minima in the [$\alpha$/Fe] histograms. For the 
intermediate metallicity stars just plotting the  [$\alpha$/Fe] histogram will not reveal the minima, because the stars
at these metallicities show a decrease of [$\alpha$/Fe] with [Fe/H] (see Fig.~\ref{fig_alfe_app}). Thus, we first detrended the 
[$\alpha$/Fe] by applying a liner fit and subtracting it. Then in the [$\alpha$/Fe] histogram we identified the minima and by adding it
to the previously applied liner fit we obtained the line which separates the high- and low-$\alpha$ stars at 
-0.3 $\leq$ [Fe/H] $\leq$ 0.0 dex. The separation lines for each metallicity bin presented in  Fig.~\ref{fig_alfe_app}.

\begin{figure}
\begin{center}
\begin{tabular}{c}
\includegraphics[angle=0,width=1\linewidth]{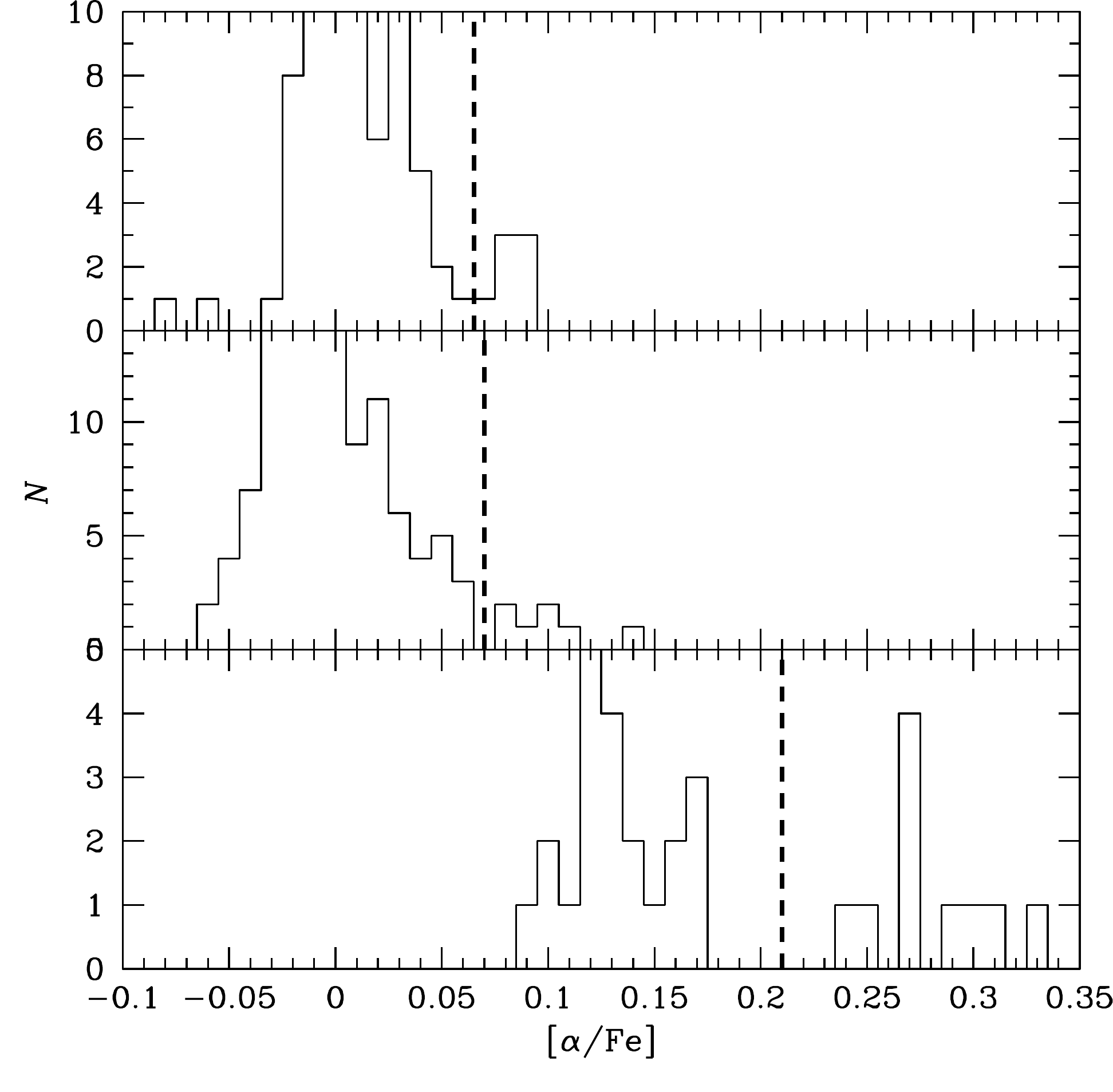}
\end{tabular}
\end{center}
\vspace{-0.6cm}
\caption{High-$\alpha$ and low-$\alpha$ separation histograms for the stars with metallicities $<$ -0.3 dex ({\it bottom}), 
-0.3 $\leq$ [Fe/H] $\leq$ 0.0 dex ({\it middle}), and [Fe/H] $>$ 0.0 dex ({\it top}). The dotted lines are the separation curves between the thin and thick disks.}
\label{fig_alfe_app}
\end{figure}

\label{lastpage}

\end{document}